\documentclass[nonacm,acmsmall,screen]{acmart}

\usepackage{soul}             
\usepackage{graphicx}         
\usepackage[inline]{enumitem} 
\usepackage{todonotes}        
\usepackage{xspace}           
\usepackage{xspace}

\usepackage[russian,english]{babel} 

\usepackage[most]{tcolorbox}  \newtcolorbox{summarybox}{
    colback=gray!5,      colframe=gray!50,    boxrule=0.3pt,       arc=2pt,             left=6pt, right=6pt, top=4pt, bottom=4pt
}

\usepackage{booktabs} 
\usepackage{multirow} 
\usepackage{makecell} 

\usepackage{array}   
\usepackage{siunitx} \ExplSyntaxOn
\cs_if_exist:NF \qty
  {
    \NewDocumentCommand \qty { O{} m m }
      {
        \SI[#1]{#2}{#3}
      }
  }
\ExplSyntaxOff
\usepackage{stmaryrd}

\usepackage{tikz}
\usetikzlibrary{external}
\tikzset{external/system call={
        lualatex -shell-escape -halt-on-error -interaction=batchmode
        -jobname "\image" "\texsource"
    }
}
\tikzexternaldisable
\usetikzlibrary{fit}
\usepackage{pgfplots}
\usepgfplotslibrary{groupplots}
\usepgfplotslibrary{colormaps}

 \newcommand{\partparagraph}[1]{\vspace{1mm}\noindent\textbf{#1:}}

\newcommand{\langFull}[0]{Signal-Spectrum Temporal Logic\xspace}
\newcommand{\lang}[0]{S2TL\xspace}

\newcommand{\funof}[1]{\left(#1\right)}           
\newcommand{\R}[0]{\mathbb{R}}                    
\newcommand{\B}[0]{\mathbb{B}}                    
\newcommand{\T}[0]{\mathbb{T}}                    
\newcommand{\F}[0]{\mathbb{F}}                    
\newcommand{\fdom}{\F}                            
\newcommand{\ftr}{\mathcal{F}}                    
\newcommand{\tftr}{\mathcal{W}}                   
\newcommand{\CPLX}[0]{\mathbb{C}}                  
\newcommand{\abs}[1]{\text{abs}\funof{#1}}
\renewcommand{\arg}[1]{\text{arg}\funof{#1}}      
\newcommand{\freqt}[1]{\text{\foreignlanguage{russian}{#1}}} 
\newcommand{\freqparenthesis}[1]{
    \mathopen{\vcenter{\hbox{\scalebox{1.1}[1.3]{$\llparenthesis$}}}}
    #1
    \mathclose{\vcenter{\hbox{\scalebox{1.1}[1.3]{$\rrparenthesis$}}}}
}

\newcommand{\freqp}[3]{\freqparenthesis{#1}^{[#2,#3]}}

\definecolor{ref}{RGB}{0,0,0} 
\definecolor{exp}{RGB}{0,128,255} 
\definecolor{act}{RGB}{255,51,253}

\title{On the Time and Frequency Domain Representations of Signals for CPS Specification}

\author{Claudio Mandrioli}
\orcid{0000-0002-7013-1191}
\email{claudio.mandrioli@uni.lu}
\affiliation{
    \institution{University of Luxembourg}
    \city{Luxembourg}
    \country{Luxembourg}
}

\author{Drishti Yadav}
\orcid{0000-0002-2974-0323}
\email{drishti.yadav@uni.lu}
\affiliation{
     \institution{University of Luxembourg}
     \city{Luxembourg}
     \country{Luxembourg}
}

\author{Domenico Bianculli}
\orcid{0000-0002-4854-685X}
\email{domenico.bianculli@uni.lu}
\affiliation{
    \institution{University of Luxembourg}
    \city{Luxembourg}
    \country{Luxembourg}
}

\begin{abstract}
Specification languages are instrumental to the Verification \& Validation of Cyber-Physical Systems (CPSs).
Most state-of-the-art specification languages use the time-domain representation of signals, which is not always suitable for describing signal shapes and dynamic behaviours.
Instead, fields like control and robotics use the frequency-domain representation to characterise these behaviours.
Time-frequency representations combine the capabilities of both domains.
We investigate the use of time-frequency representations to specify CPS requirements.
We analyse existing taxonomies of CPS requirements to identify which requirement classes can benefit from time-frequency representations.
We derive the desiderata for a specification language that uses time-frequency representations and propose \langFull (\lang), a language enabling assertions over frequency intervals and relations between frequency components.
We operationalise the \lang semantics for monitoring CPS traces, and implement a monitor.
We define specification templates for the identified requirement classes and compare time- and time-frequency-domain formulations in terms of applicability, expression fidelity, and noise tolerance of monitoring.
We observe that, while time-domain specifications are applicable only to input traces containing step-like changes or using constant interpolation, time-frequency specifications extend their evaluation to traces generated through linear interpolation, and improve tolerance to offset and high-frequency noise, while achieving comparable fidelity to the intended system properties.
\end{abstract}


 
\begin{document}

    \maketitle
    \section{Introduction}
\label{sec:introduction}
Cyber-Physical Systems (CPSs) are characterised by the tight
interaction of software and physical components and are pervasive in
our daily lives~\cite{song2017};
common examples of CPSs include robots, aircraft, drones, electric grids, smart medical devices, and satellites.
Like these examples, many CPSs are safety-critical; therefore, their rigorous Verification \& Validation (V\&V) is of paramount importance.
However, the V\&V of CPSs is a time-consuming and resource-intensive process~\cite{Garcia:2020}.
Formal specification languages support the V\&V process by enabling the precise formalisation of requirements.
Then, using monitors, the specified requirements can be automatically evaluated on execution traces to assess whether the traces satisfy the requirements.
In this way, specification languages and the associated monitor can improve both the rigour and efficiency of various V\&V activities such as runtime verification, testing, and falsification~\cite{Bartocci:2018}.
Their effectiveness, however, depends on their ability to capture the properties of interest while abstracting away contingent variations of individual executions.
Intuitively, making strong assumptions on the execution traces limits the applicability of a specification to diverse scenarios.
Analogously, if the monitoring is easily perturbed by small variations in the signals, it provides limited support for systematic and automated V\&V.

At a high level, the objective of a CPS is to steer the physical component towards a desired state.
For example, in a drone, the software uses accelerometer and GPS measurements to compute the propellers commands and fly the drone towards a target position, i.e., the desired state.
The evolution of both the actual and desired physical states can be represented as time-varying signals.
Accordingly, CPS requirements describe how the actual-state signal should evolve with respect to the desired-state (reference) signal.
This concerns not only the difference in value between the two signals, but also their shape and relation over time.
For example, requirements specify how close the drone should get to a target position, but also how fast it should reach it and whether it may overshoot while doing so.
Common examples of specified behaviours include oscillations, overshoots, settling time, or steady-state tracking~\cite{Kapinski:2016b,Chaima:2021}.

Formal specification languages are languages with a formal syntax and semantics that allow for the rigorous specification of requirements.
Once a requirement is specified, it can then be checked using runtime
verification techniques~\cite{Bartocci2018} on system traces (e.g., signals recording in the case of CPSs), providing a verdict of whether the requirement specification is satisfied or not.
State-of-the-art specification languages for CPSs, including Signal Temporal Logic~\cite{Maler:2004} (STL) and its variants~\cite{Brim:2014,Menghi:2021}, are based on the time-domain representation of signals.
That is, they define assertions on the sequence of values that signals take over time.
This representation is well suited for expressing requirements that
directly constrain signal values, such as an upper bound on a signal value.
However, it is less natural for expressing requirements that concern signal shapes, such as an oscillation, or the dynamic relation between multiple signals, such as an overshoot.
In fact, capturing such shape-related properties in the time domain requires explicitly constraining signal values at multiple time instants, and encoding assumptions about specific signal patterns and expected signal evolution.
For example, describing a sinusoidal oscillation requires constraining the signal values within one oscillation period so that they are sufficiently close to shape of the sinusoid of interest; describing an overshoot requires first encoding the shape of the reference value change (e.g., a step-like change) and then constraining the value of the actual-state signal within a time window around the reference change.
In such time-domain descriptions, the resulting specifications are often tightly coupled to particular signal realisations which makes them sensitive to contingent variations.

The frequency-domain representation of signals, already used in related fields such as control theory and robotics~\cite{Astrom:2008}, can help mitigate these limitations.
In the frequency domain, a signal is represented as a combination of sinusoidal components with different frequencies, amplitudes, and phases, called spectrum.
Rather than describing the values assumed by the signal at individual time instants, this representation captures the rates at which the signal varies.
Intuitively, slowly varying signals are dominated by low-frequency components, whereas rapidly varying signals contain more pronounced high-frequency components.
As a result, the frequency domain is particularly well suited to expressing shape-related properties.
For example, a sinusoidal oscillation appears as a isolated frequency component (a spike in the spectrum), or an overshoot appears as the amplitude amplification of a frequency component from the input to the output.

However, in a purely frequency-domain representation, the temporal information is implicitly encoded and thus ``hidden'' in the sinusoids.
As a result, it becomes difficult to locate signal features in time and assert whether a particular property holds within the signal at a certain point in time.
To address this limitation and combine the benefits of both the time and frequency domains, several hybrid time-frequency representations have been proposed~\cite{Boashash:2016}.
In a time-frequency representation, a signal is represented as a sequence of frequency components whose amplitudes and phases evolve over time.
This representation enables reasoning jointly about the frequency-based shape-oriented features and their appearance at different points in time.

The central hypothesis of this paper is that time-frequency representations provide a more suitable foundation for specifying CPS requirements on signal shapes.
Using the time-frequency domain offers two potential benefits.
First, it enables specifications to abstract from individual signal values and capture shape-related features more directly while preserving their location in time, thereby increasing their applicability across different signal shapes.
Second, it improves the tolerance of monitoring algorithms to signal noise by enabling the isolation of the frequency components that encode the relevant system behaviour.

Despite the widespread use of the frequency domain in control and robotics, its potential use in CPS specification languages remains under-explored.
Existing specification languages are almost exclusively formulated over time-domain signals and therefore provide limited support for reasoning about signal shape independently of specific signal values.
The only exception is Time-Frequency Logic (TFL)~\cite{donze:2012}, which introduces assertions over time-frequency representations of signals.
However, as the authors themselves acknowledge in their conclusions,
TFL is largely time-oriented, and, to the best of our knowledge, it has
been applied (through an extension called PTFL) to CPSs in only two subsequent
works~\cite{Nguyen:2017,Beg:2021} to perform anomaly detection.
Consequently, there is currently little understanding of 
\begin{enumerate*}[label=(\roman*)]
    \item which CPS requirements actually benefit from time-frequency representations,
    \item which language constructs are needed to express them, and
    \item whether such representations provide practical advantages for specification assessment.
\end{enumerate*}

To address these questions, in this work, we first analyse existing taxonomies of signal-based CPS requirements~\cite{Kapinski:2016b,Chaima:2021} to identify which CPS requirement classes can benefit from the use of the time-frequency representation.
We observe that requirements concerning signal features and step-response behaviours can be naturally reformulated in terms of frequency-domain properties.
Based on this observation, we then derive a set of desiderata for a specification language using time-frequency representations of signals.
Notably, we contend that such a language shall support predicates over frequency intervals and express logical relations between frequency components of different signals.

We observe that TFL does not satisfy these desiderata.
Specifically, TFL predicates on individual frequency components, and lacks constructs for expressing logical relations over frequency intervals and between frequency components of different signals.
To address this limitation, we propose \langFull (\lang, read as ``S-quare TL'').
\lang enables predicates over frequency intervals and the formulation of logical relations between frequency components of multiple signals.
Using \lang, we propose templates for the classes of CPS requirements
that can benefit from time-frequency reasoning and compare them
against corresponding STL formulations.
We then operationalise the proposed semantics and implement a
discrete-time offline monitoring algorithm for \lang as an extension
of the RTAMT monitoring tool~\cite{yamaguchi2024}.

We use our \lang templates and STL templates from the literature~\cite{Kapinski:2016b} to formulate requirements for a drone and an aircraft.
We then evaluate the resulting specifications on both realistic CPS traces and synthetic traces for which the ground-truth satisfaction of the requirements is known.
We evaluate the resulting specifications along three complementary dimensions.
First, we investigate their applicability, namely the set of traces for which the requirements can be meaningfully assessed.
Second, we assess expression fidelity, that is, how well the specifications capture the intended property.
To enable this assessment of the expression fidelity, we use control theory to generate traces that intrinsically satisfy or violate the intended property. 
Third, we assess the noise tolerance of the specification monitoring.

Our results show that, while time-domain specifications are applicable only to traces that contain step-like input changes or use constant interpolation between control points, time-frequency-based specifications can be applied also to traces generated through linear interpolation.
This applicability extension is particularly important because linear interpolation is the standard choice in most CPS applications.
For traces to which both representations are applicable, we observe comparable ability to capture the intended requirements, although with different trade-offs between soundness and precision.
Finally, time-frequency-based specifications exhibit substantially greater tolerance to offset and high-frequency noise, while achieving comparable tolerance to white noise.

To summarise, our contributions are the following:
\begin{itemize}
    \item An analysis of signal-based CPS requirement classes, identifying those that benefit from the time-frequency representation of signals.
    Based on this, we derive the desiderata for a specification language able to express them.
    \item The formal definition of \lang language in terms of its
      syntax and semantics, which enables time-frequency specifications with predicates over frequency intervals and logical relationships between frequency components.
    \item A set of specification templates for expressing signal-based CPS requirement classes.
    \item An empirical comparison---in terms of applicability,
      expression fidelity, and noise tolerance of monitoring---of time-frequency specifications expressed in \lang\ and time-domain specifications expressed using STL templates from~\citet{Kapinski:2016b}.
\end{itemize}

The rest of the paper is structured as follows.
Section~\ref{sec:background} introduces the relevant background on signal-based CPS requirements, as well as time-, frequency-, and time-frequency-domain signal representations.
Section~\ref{sec:properties} discusses which CPS requirement classes can benefit from the use of the time-frequency domain and qualitatively explains how these requirements can be captured in this domain.
Section~\ref{sec:syntax-and-semantics} introduces the formal syntax and semantics of the \lang language.
Section~\ref{sec:templates} presents our \lang templates for formulating the different requirement classes.
Section~\ref{sec:evaluation} reports on our empirical comparison between time-domain and time-frequency-domain specifications.
Section~\ref{sec:outlook} discusses the implications of this work for software engineering researchers and practitioners.
Section~\ref{sec:related} discusses the related work.
Finally, Section~\ref{sec:conclusions} concludes the paper.

     \section{Background}
\label{sec:background}
In this section, we present the background on signal-based CPS requirements and their specification.
First, we present an overview of the requirement classes based on two taxonomies from the literature~\cite{Chaima:2021,Kapinski:2016b}.
Second, we describe how signals are represented in the time, frequency, and time-frequency domains.
Finally, we briefly illustrate STL and TFL, two state-of-the-art specification languages for expressing CPS requirements.

\begin{table}
    \centering
    \caption{Classes of signal-based CPS requirements proposed in the literature (left and centre columns), and their harmonization proposed in this paper (right column).}
    \label{tab:cps-requirements}
    \begin{tabular}{llll}
\toprule
         & {\bf \citet{Chaima:2021}}       & {\bf \citet{Kapinski:2016b}}         & {\bf Harmonised}       \\ \midrule
{\bf UA} & Data Assertion (Untimed)        &                                      & Untimed Data Assertion \\ \hline
{\bf TA} & Data Assertion (Timed)          &                                      & Timed Data Assertion   \\ \hline
{\bf FR} & Functional Relation             &                                      & Functional Relation    \\ \hline
{\bf TR} & Order Relation                  & Timed Relation                       & Timed Relation         \\ \hline
{\bf SP} & Spikes                          & Spikes (and glitches)                & Spikes                 \\ \hline
{\bf OS} & Oscillation (damped/driven)  & Ringing                              & Oscillation            \\ \hline
{\bf SS} &                                 & Steady-State Error                   & Steady-State Error     \\ \hline
{\bf RT} & Rise/Fall Time                  & Rise/Fall Time                       & Rise/Fall Time         \\ \hline
{\bf OV} & Over/Under-shoot                & Over/Under-shoot                     & Over/Under-shoot       \\ \hline
{\bf ST} &                                 & Settling Time                        & Settling Time          \\
\bottomrule
\end{tabular}
 \end{table}

\subsection{Signal-Based CPS Requirements}
\label{sec:background-requirements}
In the literature, two taxonomies of signal-based CPS requirements have been proposed based on applications in the automotive~\cite{Kapinski:2016b} and aerospace~\cite{Chaima:2021} domains.
Table~\ref{tab:cps-requirements} lists the classes identified in the two works (left and centre columns).
Despite the different context, the two taxonomies identify very similar classes, supporting their general validity across application domains.
In this paper, we propose a ``harmonised'' classification (right column), with the nomenclature intended to unify previous taxonomies.
Below, we briefly describe each class and give a representative example in the context of an aircraft control system.

\noindent{\bf Untimed Data Assertion (UA)}:
It specifies a constraint on the signal values that is expected to
hold over time.
{\em Example:} the desired altitude sent to the plane autopilot must always be at least \qty{1000}{\metre} above ground level.

\noindent{\bf Timed Data Assertion (TA)}:
It specifies a constraint on signal values within specific time intervals.
{\em Example:} the plane actual altitude shall remain above \qty{1000}{\metre} from ground level from three minutes after takeoff until two minutes before landing.

\noindent{\bf Functional Relation (FR)}:
It captures a mathematical relation between two or more signals.
{\em Example:}  the fuel consumption of a plane shall be quadratic
with the plane speed.

\noindent{\bf Timed Relation (TR)}:
It expresses precedence or response between events, values or signal patterns, possibly with constraints on the temporal distance of precedence or response.
{\em Example:}  if an aircraft pilot activates the autopilot, it must take full control of the aircraft within \qty{10}{\second}.

\noindent{\bf Spike (SP)}:
It is a sudden, short-lived and large change in signal values, typically undesired.
{\em Example:}  no spikes with an amplitude exceeding \qty{100}{\volt}
shall be sent to actuators.

\noindent{\bf Oscillation (OS)}:
It is the repetition of a pattern (e.g., sinusoidal or triangular) over a time period and around an average value.
An oscillation may be damped (decreasing in amplitude) or driven (increasing in amplitude).
{\em Example:} the plane yaw angle (informally, the direction of the plane ``nose'') shall not exhibit oscillations with a period lower than \qty{3}{\second} for more than \qty{10}{\second}.

\noindent{\bf Steady-State error (SS)}:
It defines an upper bound of the steady-state error value, that is the difference between the desired value of a physical quantity and its actual one when the system is at steady state, i.e., when the two values have been constant for a while.
{\em Example:}  a plane cruising at constant altitude shall be at least within \qty{10}{\metre} from the desired altitude.

\noindent{\bf Rise/Fall Time (RT)}:
It prescribes the time within which a physical signal should reach a value close to the desired one after a step-like change in the latter.
It is usually defined as the time required for a signal to rise (or fall) from 10\% to 90\% of the step size~\cite{Astrom:2008}.
{\em Example:} when a pilot issues a change in plane altitude from \qty{8000}{\metre} to \qty{10000}{\metre}, the plane shall go from \qty{8200}{\metre} to \qty{9800}{\metre} within \qty{180}{\second}.

\noindent{\bf Overshoot (OV)}:
It prescribes how much the actual value can exceed the new desired value, after a step-like change in the desired value.
{\em Example:} using the same scenario as for RT, the maximum accepted overshoot might be \qty{20}{\metre}, meaning that, during the altitude change, the plane shall not go above \qty{10020}{\metre}.

\noindent{\bf Settling Time (ST)}:
It constrains the time it takes for a signal to enter and remain within a predefined range around its desired value after a step-like change.
{\em Example:} using again the plane altitude example, it would be the time it takes for the plane to reach and stay within $10000\pm\qty{10}{\metre}$.

\begin{figure}
    \centering
    \includegraphics{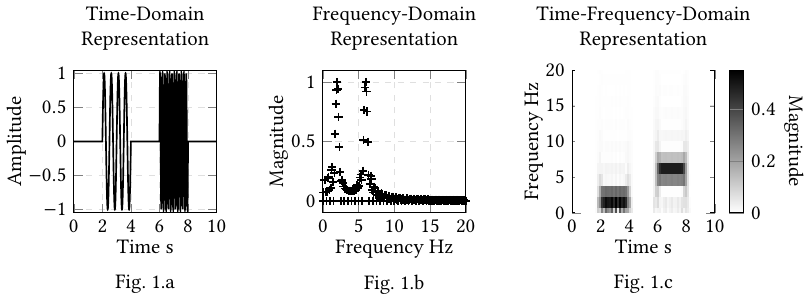}
    \caption{Illustration of how a signal is represented in the time, frequency, and time-frequency domains.
    }
    \label{fig:tf-representation-examples}
\end{figure}

\subsection{Time-, Frequency-, and Time-Frequency-Domain Representation of Signals}
\label{sec:background-tf-primer}
\subsubsection{Time-Domain Representation of Signals}
\label{sec:background-time}
In the time domain, a signal is represented as a sequence of values over time~\cite{Maler:2004}.
Formally, given a time domain $\T$, a signal $s$ is a function $s:\T\to\R$, where $\R$ is the set of real numbers.
In this paper, we consider discrete-time signals, sampled at regular time intervals, and defined over a finite domain.\footnote{
    The assumption of a finite time domain is generally acceptable as most CPS applications have a time window within which either the requirements are satisfied, or they will never be.
}
This means that $\T$ is a finite set $\{t_0,t_1\dots t_{\max}\}$ of equally-spaced time values.
The spacing $\mathit{dt}=t_i-t_{i-1}$ is called \emph{sampling time}.
In Figure~\ref{fig:tf-representation-examples}.a, we show the time-domain representation of a signal that exhibits two temporary oscillations, the first with a period of \qty{0.5}{\second} in the interval $[2\unit{\second},4\unit{\second}]$ and the second with a period of \qty{0.166}{\second} in the interval $[6\unit{\second},8\unit{\second}]$, and that otherwise remains constant.
As usual, the time-domain representation associates a signal value with each point in time.

\subsubsection{Frequency-Domain Representation of Signals}
\label{sec:background-freq}
In the frequency domain, a signal is represented as the sum of sinusoids of different frequencies.\footnote{
    This section covers the fundamentals of Fourier analysis within the limits of what is needed for this work.
    A more complete discussion of Fourier analysis can be found in~\citet{Bracewell:2000}.
}
For each sinusoid\textemdash also called {\em frequency component}\textemdash this representation defines an amplitude and a phase, which can be represented as the modulus and phase of a complex number.
This sequence of amplitudes and phases is called the signal's {\em frequency spectrum} (or simply the spectrum), and is obtained from the time-domain representation using the Fourier transform $\ftr[\cdot]$.\footnote{
    We use the square brackets to distinguish transforms from functions.
}
Formally, given a frequency domain $\fdom$, the Fourier transform of a signal $s$ is a function $\ftr[s]:\F\to\CPLX$, where $\CPLX$ is the set of complex numbers.
In the remainder of this paper, given the greater importance of the amplitude part of the spectrum compared to the phase part, when discussing the spectrum or specific frequency components of a signal in either the frequency or the time-frequency representation, we refer to their sole amplitudes, unless stated otherwise.
In the case of discrete-time signals considered in this work, the frequency domain $\F$ is likewise discretised with equally spaced frequency values $\{0,f_1,f_2,\dots f_{\max}\}$.
The spacing $\mathit{df}=f_i-f_{i-1}$ is the \emph{frequency resolution}.

When computing a spectrum using the Fourier transform, the sampling of the frequency axis (i.e., the frequency components used to represent the signal) is directly related to the sampling of the time axis of the original signal.
Intuitively, a longer signal provides a better observation of slowly varying components.
Formally, a longer signal (i.e., a signal with a larger $t_{\max}$) provides a higher frequency resolution, given by $\mathit{df}=1/t_{\max}$.
Conversely, the faster (i.e., more frequently) a signal is sampled, the better we can observe fast-changing (i.e., high-frequency) components.
Specifically, the sampling interval $\mathit{dt}$ determines the highest detectable frequency, according to the relation $f_{\max}=1/\mathit{dt}$.

At an intuitive level, the amplitude of a frequency component is a measure of how strongly the time-domain signal resembles the corresponding sinusoid.
Accordingly, a slow-changing signal exhibit larger amplitudes for low-frequency components, while a fast-changing signal exhibit larger amplitudes for high-frequency components.
This emphasis on how a signal changes, rather than on its values at individual time instants, makes the frequency domain well suited to concisely expressing properties related to signals shapes.
On the other hand, the frequency domain hides the temporal localisation of signal features: it does not indicate when particular features occur within the signal. 
Intuitively, because each frequency component is a measure of similarity between the corresponding sinusoid and the signal as a whole, it is difficult to determine which specific portion of the signal gives rise to this component.
For example, in Figure~\ref{fig:tf-representation-examples}.b we show the frequency-domain representation of the signal introduced in the left-hand side plot.
In this representation, the horizontal axis corresponds to the frequency components, and each frequency component is associated with an amplitude.
The frequency spectrum of the signal exhibits two prominent spikes at \qty{2}{\hertz} and \qty{6}{\hertz}, corresponding to the inverse of the periods of the two oscillatory segments (representing sinusoids) in the signal. Although these peaks make the oscillations easy to identify, the spectrum alone does not reveal when they occur in the signal.

Finally, in the frequency domain, when considering systems that transform input signals into output signals, it is often useful to compare the spectra of the input and output. In this context, a frequency component is said to be {\em amplified} if its amplitude is greater in the output than in the input, and {\em attenuated} (or {\em filtered}) if its amplitude is smaller.

\subsubsection{Time-Frequency-Domain Representation of Signals}
\label{sec:background-time-freq}
The time-frequency domain representation of a signal seeks to combine the ability of the frequency domain to capture shape-related features of a signal with the ability of the time domain to localise such features in time.\footnote{
    The background presented in this subsection is based on~\citet{Boashash:2016}.
}
In the time-frequency domain, a signal can be seen either as a sequence of spectra over time or, equivalently, as a set of signals describing the time-varying amplitudes of individual frequency components.
This representation is obtained by applying a time-frequency transform $\tftr[\cdot]$.
Formally, given a time domain $\T$, and a frequency domain $\F$, the time-frequency representation of a signal $s$ is a function $\tftr[s]:\T\times\F\to\CPLX$.
Thus, for a fixed time $\bar{t}$, $\tftr[s]\funof{\bar{t},f}$ is a frequency spectrum, whereas for a fixed frequency $\bar{f}$, $\tftr[s]\funof{t,\bar{f}}$ is a signal representing the temporal evolution of that frequency component.

There are numerous time-frequency representations of a signal, among which the most widely used are the Short-Time Fourier Transform (STFT), the Wavelet Transform (WT), and the Wigner-Ville Distribution.
The main differences between these representations concern 
\begin{enumerate*}[label=(\roman*)]
    \item their frequency and time domain resolutions (i.e., the precise identification of individual frequency components, and their precise localisation in time, respectively), and
    \item the presence of artefacts known as cross-terms (i.e., interference artefacts that appear in the time-frequency representation between or around true signal components and do not correspond to actual physical features of the signal).
\end{enumerate*}
Time and frequency resolutions are subject to an inherent trade-off arising from the Heisenberg uncertainty principle~\cite{Cohen:1995}.
This means that, even theoretically, it is impossible to achieve arbitrarily high resolution in both domains simultaneously.
Moreover, methods that provide high time-frequency resolution produce cross-terms.
The presence of such terms may complicate the interpretation of the time-frequency representation and thus introduces an additional practical trade-off between their suppression and the achievable time and frequency resolutions.
As a result, the choice of the appropriate time-frequency representation (precisely, the transform) is application-specific.

In the remainder of this paper, we employ the STFT because it does not
generate cross-terms, and is conceptually the closest extension of the
Fourier transform; these properties make the STFT particularly
straightforward to interpret as compared to other
transforms.\footnote{It is important to note that alternative time-frequency representations can provide higher resolution when their associated trade-offs are acceptable.}
The STFT is based on the idea of ``windowing'' a signal with a kernel or window function to extract a local signal portion around a given point in time, and then applying the Fourier transform to obtain the local spectrum.
The \emph{length} and \emph{shape} of the window function determine the resulting time-frequency resolution: a short window function provides higher time resolution, while a longer window provides higher frequency resolution.
Further, the shape of the window determines how the signal samples are weighted and also influences the time-frequency resolution~\cite{oppenheim:2010}.
Common choices include rectangular and triangular windows:
a rectangular window assigns equal weights to all the signal samples within the window, whereas a triangular window assigns greater weight to samples near its centre and linearly decreasing weight towards its edges.
In general, a rectangular window offers better frequency resolution but is more prone to interference from nearby frequency components;
conversely, a triangular window suppresses such interference more effectively, although this comes at the cost of reduced frequency resolution.

Figure~\ref{fig:tf-representation-examples}.c shows the time-frequency representation of the same signal as in the other two plots.
In this example, we use the STFT with a triangular window of length \qty{1}{\second}.
Since this representation is a function of both time and frequency, it
is displayed as a two-dimensional colour map:
the horizontal axis represents time, the vertical axis represents frequency, and the colour (according to the bar on the right) indicates the intensity of each frequency component at each point in time.
From this plot, we can see that the time-frequency representation combines the benefits of both the time and frequency domains.
In fact, it exhibits a first high-intensity (higher amplitudes) region in the time interval $[2\unit{\second},4\unit{\second}]$ at \qty{2}{\hertz}, and a second one in the interval $[6\unit{\second},8\unit{\second}]$ at \qty{6}{\hertz}.
These regions correspond to the two oscillatory segments of the original signal, illustrating how the time-frequency representation simultaneously captures both the signal's frequency content and time-locality of when the frequency components are present.

\subsection{STL and TFL}
\label{sec:background-langs}

\subsubsection{STL (Signal Temporal Logic)}
\label{sec:stl}
The standard specification language for CPS is STL (Signal Temporal Logic)~\cite{Maler:2004}.
STL is based on Metric-Interval Temporal Logic (MITL)~\cite{Henzinger:1996}, a decidable fragment of Metric Temporal Logic (MTL)~\cite{Koymans:1990}. MTL extends Linear Temporal Logic (LTL)~\cite{Pnueli:1977} with quantitative timing constraints.
The syntax of MITL is:
\begin{equation}
    \varphi ::= p | \neg\varphi | \varphi_1\lor\varphi_2 | \varphi_1\mathcal{U}_{[a,b]}\varphi_2
\end{equation}
where $p$ is a Boolean predicate, $\neg$ and $\lor$ denote logical negation and disjunction, respectively, and $\mathcal{U}_{[a,b]}$ is the time-bounded until operator.
STL extends MITL with predicates over signal values $\mu_1\funof{x},\dots\mu_n\funof{x}$, with $\mu_i:\R\rightarrow\B$, which constitute the atomic propositions that the language builds upon.
In practice, these predicates are typically inequalities involving signal values.
During monitoring, the signal predicates can be used to translate signals into Boolean predicates, thereby allowing an STL formula to be interpreted as an MITL formula over time.
Importantly, STL is used in one of the two CPS requirement taxonomies discussed in Section~\ref{sec:background-requirements} to write formula templates capturing the different CPS requirement classes~\cite{Kapinski:2016b}.

\subsubsection{TFL (Time-Frequency Logic)}
TFL (Time-Frequency Logic) further extends STL with predicates on the time-frequency representation of signals~\cite{donze:2012}.
In practice, it introduces a set of operators that extract the amplitude of a given frequency from the STFT of a signal.
Using the notation from Section~\ref{sec:background-time-freq}, these operators compute $\tftr_{w,L}[s]\funof{t,\bar{f}}$, where $\bar{f}$ is the frequency of interest, $w$ is the window shape, and $L$ is the window length.
These operators enable a monitoring procedure analogous to that of STL.
For a given frequency $\bar{f}$, the corresponding $\tftr_{w, L}[s]\funof{t,\bar{f}}$ operator defines a new time-domain signal whose value at each time instant is the amplitude of the corresponding frequency component.
STL predicates $\mu$ can then be written on this derived signal, reducing the monitoring problem to that of evaluating the corresponding MITL formula.

The original TFL paper applied the logic to the problem of identifying a musical melody within a song.
Subsequent work~\cite{Nguyen:2017} proposed a variant by replacing STFT with the WT, which offers higher time-frequency resolution, and applied it to the classification of abnormal data in a hydrogen fuel cell and an electrocardiogram.

     \section{CPS Requirements in the Frequency Domain}
\label{sec:properties}
In this section, we identify the classes of signal-based CPS requirements that can benefit from a frequency-domain representation of signals.
Drawing on results from control theory~\cite{Astrom:2008}, we describe how these requirement classes can be formalised using such representations.
We conclude by defining a list of desiderata that a specification language should fulfill to support predicates over the time-frequency representation of signals.

\subsection{Requirements Classes that Can Benefit from Using the Frequency-Domain}
\label{sec:partitioning-req-classes}
Among the requirement classes listed in
Table~\ref{tab:cps-requirements}, SP, OS, SS, RT, OV, and ST can benefit from the frequency-domain representation of signals.
All of these classes define assertions about signal shapes, either for a single signal (SP and OS) or for the input and output signals of a CPS (SS, RT, OV, and ST).
Specifically, SP and OS describe {\em features} that can appear in a signal shape.
Conversely, SS, RT, OV, and ST describe the expected shape of the response of a physical quantity to a step-like change in its corresponding desired value.
In other words, these requirements express an {\em implication} between a precondition on the set-point signal (namely, the presence of a step-like change) and a postcondition on the signal of the actual physical quantity.
In the remainder of this paper, we refer to the SP and OS classes as the {\bf signal-feature requirement classes}; and to the SS, RT, OV, and ST classes as the {\bf step-response requirement classes}.

In contrast, the remaining requirement classes (UA, TA, TR, and FR) generally may not benefit from a frequency-domain representation of signals.
The main reason is that they specify properties directly in terms of signal values over time.
Specifically, the UA and TA requirement classes define constraints on signal values.
In practice, such constraints typically take the form of inequalities that must hold either at every point in time (for UA) or only within specific time intervals (for TA).
Because these requirements directly define assertions on the time-domain values of a signal, they are most naturally specified using the time-domain representation of the signal.

Among the remaining requirement classes, FR requirements can be seen as equality constraints between two signals, where one signal is generated by applying a function to another signal.
For example, if the specification states that an aircraft's fuel consumption should be quadratic to the speed, then the requirement is that the fuel consumption signal should be equal to the signal obtained by squaring the speed signal.
In practice, the equality between the two signals can be expressed as an inequality that constrains the difference between their values to lie within a prescribed tolerance.
Because this comparison is performed directly on the signal values,
the time-domain representation would be the most natural choice.
In some specific cases, however, it may be beneficial to compare the signals in the frequency domain, for example to filter-out high-frequency noise or to remove low-frequency trends (i.e., to ignore differences in their average values).
Since the usefulness of such an approach is application-dependent, we do not consider a frequency-domain specification of the FR requirement class in the scope of this work.\footnote{
    A frequency-domain formulation of FR requirements could be
    implemented using an approach similar to the one described in \S\ref{sec:qualitative-reqs-f-formulation} for SS requirements, where only selected frequency ranges of the two signals are compared.
}

Finally, TR requirements specify temporal relationships between events, signal values or signal patterns, and are therefore closely tied to the sequentiality of what they assert over time.
In some applications, the ``events and signal patterns'' may be best captured using a frequency-domain representation.
However, this affects only the definition of the atomic predicates, not the temporal relationships between them. Accordingly, we consider TR requirements to be most naturally expressed using the time-domain representation of signals.

\subsection{Frequency-Domain Formulation of CPS Requirements}
\label{sec:qualitative-reqs-f-formulation}
\begin{figure}[t]
    \centering
    \includegraphics{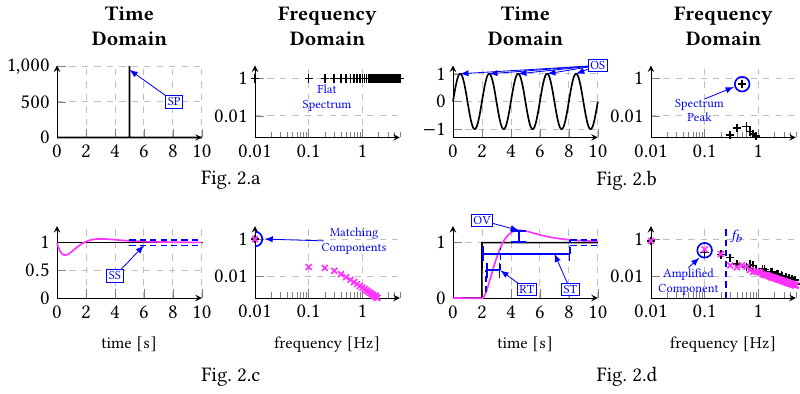}
    \caption{Illustration of the SP, OS, SS, OV, RT, and ST requirements in time and frequency domain.
    }
    \label{fig:freq-specs-examples}
\end{figure}

To illustrate how the requirements translate into the frequency domain, we use the representative signal traces in Figure~\ref{fig:freq-specs-examples}.
This figure shows four signals\textemdash a spike (Figure~\ref{fig:freq-specs-examples}.a), a sinusoid (Figure~\ref{fig:freq-specs-examples}.b), a CPS constant-input response (Figure~\ref{fig:freq-specs-examples}.c), and a CPS step-input response (Figure~\ref{fig:freq-specs-examples}.d)\textemdash each displayed in both the time (left) and frequency (right) domains.
For the input-response requirements classes
(Figure~\ref{fig:freq-specs-examples}.c and
Figure~\ref{fig:freq-specs-examples}.d), the input signal is shown in
black and the output signal in purple. The signal features relevant to
assessing the requirements are highlighted in blue in both domains.

\partparagraph{SP}
The defining feature of a spike is its sudden and large change in signal value.
Such a rapid variation corresponds to large high-frequency components in the frequency domain.
As shown in the right-hand plot of Figure~\ref{fig:freq-specs-examples}.a, a spike exhibits a nearly ``flat spectrum'', with the same amplitude across the frequency range, including the higher frequencies.
Consequently, a requirement specifying the absence (or presence) of spikes can be reformulated in the frequency domain as an upper (or lower) bound on the amplitudes of the high-frequency components.
Importantly, in the frequency-domain formulation, the requirement constrains any signal with large, rapid changes.
This broader interpretation is often a desirable extension, as large and rapid signal changes can be harmful for hardware components or characterise sensor faults.

\partparagraph{OS}
The frequency domain is well suited for identifying repeated signal patterns, as it is based on the representation of signals as sum of sinusoids, which are inherently periodic.
As illustrated in Figure~\ref{fig:freq-specs-examples}.b, a purely sinusoidal signal is translated to a sharp peak in the frequency spectrum.
In general, periodic repetitions of any shape in the time-domain translate to sharp peaks in the frequency spectrum.
Thus, assertions on the presence of oscillations in the time domain
can be reformulated in the frequency domain as assertions on the presence of sharp peaks in the spectrum.
Notably, detecting oscillations directly in the time domain typically requires comparing different parts of the signal or identifying recurring sequences of peaks and dips, making the analysis less straightforward.

\partparagraph{SS}
A signal is in steady state when it maintains a constant value, typically after an initial transient (e.g., after a step-like change).
In the left-hand plot of Figure~\ref{fig:freq-specs-examples}.c, the desired-state signal (black) is constant, and therefore remains in steady state throughout the observation period.
In contrast, the actual-state signal (purple) reaches steady state only after \qty{5}{\second}.
From that point onward, the two signals are sufficiently close to satisfy the SS requirement, as shown by the dashed blue line.
In the frequency domain, the constant desired-state signal contains only a low-frequency component (the component of amplitude 1 at \qty{0.01}{\hertz}), whereas the actual-state signal contains also some higher-frequency components which correspond to the initial transient oscillations.
Therefore, to compare only the steady-state values and ignore the initial oscillations, we can compare the sole \qty{0.01}{\hertz} components of the two signals, and disregard the higher-frequency ones, as highlighted in the right-hand plot of Figure~\ref{fig:freq-specs-examples}.c by the SS blue arrow.
More generally, the SS requirements can be reformulated in the frequency domain as the matching of low-frequency components of the desired- and actual-state signals.
Conversely, evaluating the same requirement in the time domain would require first distinguishing that the actual-state signal is at steady state only after \qty{5}{\second}, and then comparing the two signals only over that interval.

\partparagraph{OV}
The OV requirement limits the extent to which the actual state can ``overreact'' to a change in the desired value.
For example, in Figure~\ref{fig:freq-specs-examples}.d, we observe that the actual-state signal (purple) temporarily exceeds the desired-state signal, as highlighted by the OV blue arrow.
In the frequency domain, this overreaction manifests as an amplification of the frequency components of the actual-state signal relative to the corresponding components of the desired-state signal.
Since such an amplification can happen during a transient, i.e., when the signals are changing, it concerns a higher frequency range.
For example, in Figure~\ref{fig:freq-specs-examples}.d, we observe that there is an amplified frequency component at \qty{0.1}{\hertz}. 
More generally, OV requirements can be defined in the frequency domain as constraints on the maximum allowed amplification between desired- and actual-state spectra.
In the time domain, such requirements would involve first the identification of the transient (e.g., of the step change in the desired state and consequent change in the actual state), then the identification of the overshoot value, and finally its comparison with the accepted amplification.

\partparagraph{RT and ST}
We discuss RT and ST together because both specify how quickly the actual state is expected to track a change in the desired state, and their frequency-domain formulations are closely related.
In the time domain, the actual-state signal can be viewed as a ``smoothed" version of the desired-state signal.
For example, in Figure~\ref{fig:freq-specs-examples}.d, the actual-state signal (purple) follows the step change of the desired-state signal (black), but with a gradual rather than abrupt transition.
In the frequency domain, this smoothing corresponds to an attenuation
of the high-frequency components (i.e., a reduction in amplitude of
the fast-changing part of the signal).
\label{signal-bandwidth}
In the spectrum shown in Figure~\ref{fig:freq-specs-examples}.d, we highlight the so-called \emph{bandwidth} $f_b$, which is the frequency threshold above which we observe the signal attenuation.
Below $f_b$, we observe instead that the frequency components of the actual-state signal approximately match or exceed those of the desired-state signal.
A large bandwidth corresponds to a fast tracking, as it indicates that higher-frequency components of the desired-state signal are found with equal or greater amplitude in the actual-state signal.
Conversely, a small bandwidth indicates slow tracking, as larger part of the input signal is attenuated or filtered.
Based on this observation, RT and ST requirements can be translated in the frequency domain as constraints on the value of $f_b$ .
At times, for ST, this bandwidth constraint may also need to be combined with a constraint on the maximum amplification, since overshoot can induce oscillations that increase the settling time.
On the contrary, in the time domain, similarly to OV requirements, ST and RT requirements involve first identifying the step-like transients, and then identifying the various thresholds that define the RT and ST (highlighted by the dashed lines in Figure~\ref{fig:freq-specs-examples}.d).

\subsection{Desiderata for a Frequency-Extended Logic}
\label{sec:desiderata}
The qualitative discussion on the frequency-domain formulation of CPS
requirements underlines that a specification language that predicates
on the time-frequency representation of a signal must allow for
\begin{enumerate}[label=(D\arabic*)]
    \item\label{des-interval} assertions on frequency intervals of a signal's spectrum, and
    \item\label{des-pred} logical predicates that relate corresponding frequency components in different signals.
\end{enumerate}
SP requirements require assertions over high-frequency ranges, SS over low-frequency ranges, and RT and ST requirements over a combination of the two to characterize the bandwidth $f_b$.
This motivates the need for assertions on frequency intervals.

For OV requirements, we instead want to reason about the amplification between the desired- and actual-state signals.
For example, we could formulate a predicate such as: ``if a frequency component is present in the desired-state signal (i.e., it has a non-zero amplitude), then the corresponding component in the actual-state signal should not have a larger amplitude.''
Such a requirement corresponds to a logical implication in the frequency dimension, i.e., an implication evaluated separately at each frequency of a given interval, and motivates the need for such logical relations.
In TFL, these predicates would require their explicit formulation for each frequency component.
Besides being tedious, such specifications are error-prone and impractical for real-world applications.
This limitation motivates the need for a new specification language that provides native support for predicates over the frequency-domain representations of signals.
Specifically, the language should enable concise expression of logical relations between corresponding frequency components of different signals and evaluate them over user-specified frequency intervals.

     \section{The \lang Specification Language}
\label{sec:syntax-and-semantics}
In this section, we introduce \lang, a specification language designed to address the limitations of the state of the art discussed in Section~\ref{sec:desiderata}.
We define \lang as an extension of STL with frequency-domain predicates.
With these predicates, we pursue two objectives.
First, we enable the encoding of logical relations between the spectra of one or more signals at a given point in time.
Second, we enable the quantification of the frequency variable within specified frequency intervals.
We first present the syntax of \lang and then introduce its qualitative (Boolean) semantics.
Finally, we present our implementation of the language as extension of the STL monitoring tool RTAMT, including some considerations about the monitoring complexity of the frequency predicates~\cite{Nickovic:2020,yamaguchi2024}.

\subsection{Syntax}
Let $\T\subset\R$ be the time domain and $\F\subset\R$ be the frequency domain, \lang is built on the following symbols and operators:
\begin{itemize}
\item A set of signal symbols $\{s_1,\dots,s_n\}$ where each $s_i$ is a function $s_i:\T\rightarrow\R$,
    \item A set of signal-value constants $C=\{c_1,\dots,c_m\}$ in $\CPLX$,
    \item A set of memoryless operators
      $O=\{+,-,\cdot,\abs{},\dots\}$ over complex numbers, and
    \item A time-frequency transform operator $\tftr[s]$, of type $\funof{\T\to\R}\to\funof{\T\times\F\to\CPLX}$.
\end{itemize}
We consider two types of values and terms: time-domain and frequency-domain ones.
We use Cyrillic letters throughout to distinguish frequency-domain values and terms from their time-domain counterpart.
Specifically, we use the $\freqt{к}$ (pronounced {\em ka}) to denote frequency-domain values and \freqt{б} (pronounced {\em be}) to denote frequency-domain terms.
\\A time-domain value term may be a constant, a signal, or a value computed by applying an operator in $O$ to one or more time-domain value terms:
\begin{flalign}
    v ::= &\quad c\,|\,s\,|\,o\funof{v_1,\dots,v_n}.
    \label{eq:time-domain-value}
\end{flalign}
A frequency-domain value term may be a constant, the time-frequency transform of a signal, or a value computed by applying an operator in $O$ to one or more frequency-domain value terms:
\begin{flalign}
    \freqt{к} ::= &\quad c\,|\,\tftr[s]\,|\,o\funof{\freqt{к}_1,\dots,\freqt{к}_n}.
    \label{eq:frequency-domain-value}
\end{flalign}
We can then define frequency-domain predicates as Boolean combinations
of inequalities over frequency-domain value terms:
\begin{flalign}
    \freqt{б} ::= \, \freqt{к}_1>_a\freqt{к}_2 \,
                | \, \freqt{к}_1>_p\freqt{к}_2 \,
                | \, \neg\freqt{б} \,
                | \, \freqt{б}_1\lor\freqt{б}_2.
    \label{eq:lang-fd-predicates}
\end{flalign}
Since the time-frequency transform of a signal is defined over the set of complex numbers, which is not naturally ordered, we define two different types of inequalities ($>_a$ and $>_p$), corresponding to comparisons based on amplitude and phase, respectively.

Finally, in a similar fashion as for STL (Section~\ref{sec:stl}), the entry points of \lang (i.e., \lang formulas) are formulas over time- and/or frequency-domain predicates, connected with Boolean operators and/or the until operator:
\begin{flalign}
    \varphi ::= \, v_1>v_2 \,
              | \, \freqp{\freqt{б}}{f_i}{f_j} \,
              | \, \neg\varphi \,
              | \, \varphi_1\lor\varphi_2 \,
              | \, \varphi_1\mathcal{U}_{[t_i,t_j]}\varphi_2\, ,
    \label{eq:lang-entry-point}
\end{flalign}
where $v$ expands based on Eq.~\ref{eq:time-domain-value} and
$\freqt{б}$ based on Eq.~\ref{eq:lang-fd-predicates}, and the time and
frequency intervals are defined in the respective domains
$t_i,t_j\in\T$ and $f_i,f_j\in\F$.
This syntax requires that every frequency predicate $\freqt{б}$ is contained in the special parentheses $\freqparenthesis{\cdot}$ with an apex containing a frequency interval $[f_i,f_j]$.
In the remainder of the paper, we assume the standard definitions of the timed globally $\square_{[t_i,t_j]}$ and eventually $\Diamond_{[t_i,t_j]}$ operators in terms of the timed until operator, and the conjunction $\land$ and implication $\Rightarrow$ in terms of the logical connectives.

\subsection{Qualitative Semantics}
A trace $\tau$ is the interpretation of the signal symbols in $\{s_1,\dots,s_n\}$ as functions over time, denoted as $\llbracket s_i\rrbracket_\tau:\T\rightarrow\R$.
The qualitative (Boolean) semantics of \lang determines whether a trace $\tau$ satisfies a formula $\varphi$ at a time instant $t\in\T$, denoted as $\funof{\tau,t}\models\varphi$ and $\funof{\tau,t}\nvDash\varphi$ otherwise.
As in other linear temporal logics~\cite{Furia:2012}, a trace $\tau$ satisfies a formula $\varphi$ if and only if $\funof{\tau,0}\models\varphi$.

Given a trace $\tau$ and a time instant $t$, time-domain value terms are interpreted as follows:
\begin{itemize}
    \item constants evaluate to their corresponding values, independently of the trace and the time instant;
    \item signal symbols evaluate to the value of the function defined by the trace $\tau$ at time $t$; and
    \item operators evaluate to the value obtained by applying the corresponding operator to the interpretations of their arguments.
\end{itemize}
Formally, we denote the interpretation of a term for a given trace
$\tau$ and time instant $t$ with $\llbracket\cdot\rrbracket_{\tau,t}$, and define it as
\begin{flalign}
    & \llbracket c \rrbracket_{\tau,t} = c \\
    & \llbracket s \rrbracket_{\tau,t} = \llbracket s \rrbracket_{\tau}\funof{t} \\
    & \llbracket o\funof{v_1,\dots,v_n} \rrbracket_{\tau,t} = o\funof{\llbracket v_1 \rrbracket_{\tau,t},\dots,\llbracket v_n \rrbracket_{\tau,t}} .
\end{flalign}
Frequency-domain values are instead interpreted, for a given trace and
time instant, as functions of a frequency variable $f$:
\begin{itemize}
    \item constants evaluate to their corresponding values, independently of the trace, the time instant, and the frequency;
    \item time-frequency transform symbols evaluate to the spectrum $\tftr[s_i]\funof{t,f}$, where $t$ is the time instant at which the value is interpreted and $f$ is a free frequency variable (that moves along the spectrum); and
    \item operators evaluate pointwise over the frequency domain, that is, for each frequency value they evaluate to the result of applying the corresponding operator to the interpretations of their arguments at that frequency.
\end{itemize}
Formally, we denote by $\llbracket\cdot\rrbracket_{\tau,t}\funof{f}:\F\rightarrow\CPLX$ the interpretation of a frequency-domain value term for a given trace $\tau$ and time instant $t$ , defined as:
\begin{flalign}
    & \llbracket c \rrbracket_{\tau,t}\funof{f}                              = c \label{eq:tf-const}\\
    & \llbracket \tftr[s] \rrbracket_{\tau,t}\funof{f}                       = \tftr_t[\llbracket s \rrbracket_{\tau}]\funof{f} \label{eq:tf-term}\\
    & \llbracket o\funof{\freqt{к}_1,\dots,\freqt{к}_n} \rrbracket_{\tau,t}\funof{f} = o\funof{\llbracket \freqt{к}_1 \rrbracket_{\tau,t}\funof{f},\dots,\llbracket \freqt{к}_n \rrbracket_{\tau,t}\funof{f}} ,
\end{flalign}
where the frequency-domain value terms in Eq.~\ref{eq:tf-term} use the ``time-curried'' version of the signal's time-frequency transform $\tftr_t[s]:\F\to\CPLX$, defined as $\tftr_t[s](f)=\tftr[s](t,f)$.

We can now define the satisfaction of the frequency-domain predicates with respect to the frequency variable $f$.
We denote with $\funof{\tau,t}\models_f \freqt{б}$ that trace $\tau$ at time instant $t$ satisfies the formula $\varphi$ for the frequency value $f$, and $\funof{\tau,t}\nvDash_f\varphi$ otherwise.
We then define the semantics of the frequency predicates as
\begin{flalign}
    &\funof{\tau,t}\models_f \freqt{к}_1>_a\freqt{к}_2
    \iff \abs{\llbracket \freqt{к}_1 \rrbracket_{\tau,t}\funof{f}}>\abs{\llbracket \freqt{к}_2 \rrbracket_{\tau,t}\funof{f}} \label{eq:tf-ineq-md}\\
    &\funof{\tau,t}\models_f \freqt{к}_1>_p\freqt{к}_2
    \iff \arg{\llbracket \freqt{к}_1 \rrbracket_{\tau,t}\funof{f}}>\arg{\llbracket \freqt{к}_2 \rrbracket_{\tau,t}\funof{f}} \label{eq:tf-ineq-ph}\\
    &\funof{\tau,t}\models_f \neg\freqt{б}
    \iff \funof{\tau,t}\nvDash_f \freqt{б} \\
    &\funof{\tau,t}\models_f \freqt{б}_1\lor\freqt{б}_2
    \iff \funof{\tau,t}\models_f\freqt{б}_1\lor\funof{\tau,t}\models_f\freqt{б}_2 ,
    \label{eq:tf-or-semantic}
\end{flalign}
where Eq.s~\ref{eq:tf-ineq-md} and~\ref{eq:tf-ineq-ph} compare respectively the amplitudes (with the function $\abs{\cdot}$) and phases (with the function $\arg{\cdot}$) of two complex values.
In the remainder of the paper, when the subscript $a$ or $p$ is omitted from a frequency domain predicate, we implicitly refer to the comparison of the amplitudes\textemdash i.e., $>$ is interpreted as $>_a$.

Finally, we define the semantics of \lang formulas as follows:
\begin{flalign}
    &\funof{\tau,t}\models v_1>v_2
    \iff \llbracket v_1 \rrbracket_{\tau,t}>\llbracket v_2 \rrbracket_{\tau,t} \\
    &\funof{\tau,t}\models \freqp{\freqt{б}}{f_i}{f_j}
    \iff \forall f\in [f_1,f_2] \funof{\tau,t}\models_f \freqt{б} \label{eq:f-quantify-qualitative}\\
    &\funof{\tau,t}\models \neg\varphi
    \iff \funof{\tau,t}\nvDash \varphi \\
    &\funof{\tau,t}\models \varphi_1\lor\varphi_2
    \iff \funof{\tau,t}\models\varphi_1\lor\funof{\tau,t}\models\varphi_2 \\
    &\funof{\tau,t}\models \varphi_1\mathcal{U}_{[t_i,t_j]}\varphi_2
    \iff \exists t' \in[t+t_i,t+t_j], \funof{\tau,t'}\models\varphi_2 \land \forall t''\in[t,t'], \funof{\tau,t''}\models \varphi_1 .
\end{flalign}
Importantly, the terms in Eq.~\ref{eq:f-quantify-qualitative} define the interpretation of the frequency variable $f$ and universally quantify it to any value in the interval $[f_1,f_2]$.
Consequently, they implement a logical {\em and} operator along the frequency axis, meaning that the frequency-domain predicate is satisfied if and only if it holds for every frequency value in the specified interval.
By the standard duality between universal and existential
quantification, existential requirements can be expressed by negating
the corresponding frequency-domain predicate.\footnote{
    Concretely, starting from the formula $\freqp{\freqt{б}}{f_i}{f_j}$, which requires $\freqt{б}$ to hold at every frequency in $[f_i,f_j]$ (the universal quantification), we can write the formula $\neg\freqp{\neg\freqt{б}}{f_i}{f_j}$ to require that $\freqt{б}$ holds for at least one frequency in $[f_i,f_j]$ (the existential quantification).
}

In summary, the evaluation of frequency predicates over frequency intervals, as defined by the semantics in Eq.~\ref{eq:f-quantify-qualitative}, implements the desideratum~\ref{des-interval} described in Section~\ref{sec:desiderata}.
The frequency-domain predicates defined in Eq.~\ref{eq:lang-fd-predicates} enable the specification of logical relations between corresponding frequency components in the the spectra of one or more signals, which fulfils desideratum~\ref{des-pred}.
For example, we can now write the example requirement mentioned in Section~\ref{sec:desiderata}\textemdash which states that, if a frequency component is present in the desired-state signal $r$, then the corresponding component in the actual-state signal $x$ should have a smaller amplitude\textemdash with the formula
\begin{equation}
    \square_{[0,t_{\max}]} \freqp{\tftr[r]>_a\mathit{th} \Rightarrow \tftr[x]<_a\tftr[r]}{f_1}{f_2}.
    \label{eq:s2tl-example}
\end{equation}
In this formula, starting from the left, the always operator $\square_{[0,t_{\max}]}$ requires the frequency predicate, i.e., the predicate contained in $\freqp{\cdot}{f_1}{f_2}$, to hold at every time step.
At a given time step $t$, the frequency predicate is satisfied if, for every frequency $f$ in the interval $[f_1,f_2]$, whenever the amplitude of a frequency component is greater than the given threshold $\mathit{th}$, i.e., $\abs{\tftr[r]\funof{t,f}}>\mathit{th}$, the corresponding output component $\tftr[x]\funof{t,f}$ is smaller, i.e., $\abs{\tftr[r]\funof{t,f}}>\abs{\tftr[x]\funof{t,f}}$.
The ability to express implications between different frequency components and evaluate them over the frequency interval $[f_1,f_2]$ showcases how \lang satisfies our desiderata.

\subsection{\lang Implementation and Monitoring Complexity}
\label{sec:lang-implementation}
To have an executable semantics of \lang, we extended RTAMT~\cite{Nickovic:2020,yamaguchi2024}, a state-of-the-art robustness monitoring tool for STL written in Python.
Specifically, we implemented a discrete-time offline \lang qualitative monitor.\footnote{
    We used the Microsoft M365 Copilot Chat powered by the GPT-5 chat model to sketch individual functions, create templates of Python modules, and debug code.
}
The discrete-time monitoring matches the application to sampled signals, while the offline monitoring circumvents the acausal (i.e., not causal) nature of the time-frequency transform (i.e., its dependency on future values of the signal).\footnote{
    The acausal nature of the time-frequency transform does not imply that there cannot be an online monitor for \lang formulas.
    However, such a monitor would incur an intrinsic delay that depends on the chosen time-frequency transform.
    We leave the study of online monitoring for \lang\textemdash or, more generally, for languages based on the time-frequency representation of signals\textemdash to future work.
}

In practice, we extended RTAMT to support the frequency predicates of Eq.~\ref{eq:f-quantify-qualitative}.
This required extending the parser to recognize the new syntax and implementing the corresponding semantics (Eqs~\ref{eq:tf-const}--\ref{eq:tf-or-semantic}).
To evaluate frequency-domain predicates, the specification object is associated with a user-provided time-frequency transform.
During monitoring, when a frequency predicate is encountered, the monitor computes the time-frequency representation of the signals, checks the predicate over the specified frequency interval and produces a Boolean verdict.
Then, the remainder of the formula can be treated as a standard STL one.

Although a thorough complexity analysis of \lang is outside the scope of this work, we observe that the cost associated with the frequency dimension depends on the chosen time-frequency transform.
Intuitively, the introduction of frequency-domain predicates adds an additional frequency dimension to the monitored data.
Intuitively, after applying a time-frequency transform, a signal of size $|\T|$ is represented over a domain of size $|\T|\cdot|\F|$.
Consequently, monitoring frequency predicates incurs additional computational and memory costs compared to standard STL monitoring.
However, for each frequency predicate, we note that the relevant frequency range is known statically from the specification (i.e., the interval $[f_1,f_2]$ in Eq.~\ref{eq:f-quantify-qualitative}).
Consequently, in our implementation, we restricted the monitoring to the frequency components within this interval, reducing the effective size of the monitored representation from $|\T|\cdot|\F|$ to $|\T|\cdot|[f_1,f_2]|$.

In practice, we found that the additional overhead introduced by the frequency dimension did not significantly affect the experiments reported in Section~\ref{sec:evaluation}.
As further discussed below, the overall computational cost was dominated by the size of the experimental campaign, namely the number of traces and formula to be monitored.

     \section{Expressing Signal-Based CPS Requirements in \lang}
\label{sec:templates}
In this section, building on the qualitative discussion in Section~\ref{sec:qualitative-reqs-f-formulation}, we use \lang to define templates for the requirement classes that benefit from the time-frequency domain representation of signals.
Throughout the formulas, we use the following naming conventions for the recurring signals and constant values:
\begin{itemize}
    \item $x$ denotes the CPS output signal in the step-response requirements, i.e., the actual physical state, or the sole signal in the signal-feature requirements;
    \item $r$ denotes the CPS input signal in the step-response requirements, i.e., the desired physical state;
    \item $e$ denotes the control error signal, defined as $e=\abs{r-x}$; and
    \item $f_b$ denotes the expected system bandwidth (defined in Section~\ref{sec:qualitative-reqs-f-formulation}, when discussing the RT and ST requirements).\end{itemize}

\subsection{Signal-Feature Requirements}
\subsubsection{Spike}
SP requirements often concern the presence of large high-frequency components in a signal.
In the context of CPS, ``high-frequency'' generally refers to frequencies above the system bandwidth.
Accordingly, we constrain the existence of components whose frequencies exceed the expected system bandwidth.
The corresponding \lang formula for detecting a spike is:
\begin{equation}
    \square_{[0,\tau]} \neg \freqp{\tftr[x]< \mathit{sp}}{f_{\min}}{f_{\max}},
\end{equation}
where
\begin{itemize}
    \item $\tau$ is the minimum duration of the spike to be considered,
    \item $\mathit{sp}$ is the amplitude threshold above which a frequency component is considered large, and
    \item $f_{\min}$, with $f_{\min}>f_b$, is determined by the maximum duration of the spike that we want to consider, as longer spikes require lower-frequency components to be detected.
\end{itemize}

\subsubsection{Oscillation}
OS requirements concern the presence of oscillatory behaviour.
In the frequency domain, an oscillation appears as a spike in the spectrum.
Accordingly, we define an assertion on the presence of an oscillation with frequency $f_o$ by requiring that at least one frequency component in the vicinity of $f_o$ exceeds a minimum amplitude threshold, while neighbouring frequency components remain sufficiently smaller.
The resulting \lang formula is:\footnote{
    The  reported formula includes only a lower bound on the amplitude of the oscillation through $\neg \freqp{\tftr[x]< \mathit{os}}{f_o-\delta}{f_o+\delta}$.
    It can be extended to include an upper bound by replacing this constraint with $\neg \freqp{\tftr[x]< \mathit{os}_{\min} \lor \tftr[x]> \mathit{os}_{\max}}{f_o-\delta}{f_o+\delta}$.
}
\begin{equation}
    \neg\freqp{\tftr[x]< \mathit{os}}{f_o-\delta}{f_o+\delta}\land
    \freqp{\tftr[x]< \nu}{f_o-\alpha}{f_o-\delta}\land
    \freqp{\tftr[x]< \nu}{f_o+\delta}{f_o+\alpha},
\end{equation}
where
\begin{itemize}
    \item $\mathit{os}$ is the minimum amplitude threshold for the oscillation,
    \item $\delta$ defines the frequency interval around $f_o$ within which the oscillation is detected,
    \item $\nu$ is the threshold below which neighbouring frequency components are considered low-amplitude components (should be set as a fraction of $\mathit{os}$), and
    \item $\alpha$, with $\alpha>\delta$, defines the frequency interval around $f_o$ in which neighbouring frequency components are required to remain below $\nu$.
\end{itemize}

\subsection{Step-Response Requirements}
\subsubsection{Steady-State Error}
SS requirements are defined as an implication, in which the precondition is that the system is in steady state (i.e., the signals are approximately constant), and the postcondition is that the desired and actual physical states match within a specified tolerance.
To express the precondition, we require that the high-frequency components of the input signal have small amplitudes.
These components capture rapid variations in the signal; therefore, when they are sufficiently small, the signal can be considered to be in steady state.
To express the postcondition, we require that the low-frequency components (i.e., the slowly varying part) of the error signal\textemdash i.e., the difference between input and output\textemdash also have small amplitudes.
We distinguish low- and high-frequency components as those below and above the expected system bandwidth $f_b$.\footnote{
    This formulation can be somewhat restrictive, as it constrains the entire frequency ranges below and above $f_b$ for the input and output signals.
    In some applications, a less restrictive assessment of this requirement may be obtained by excluding a small frequency band around $f_b$, for example by using the intervals $[f_b+\delta,f_{\max}]$ and $[0,f_b-\delta]$ instead.
}
The resulting \lang formula is:
\begin{equation}
    \square_{[0,t_{\max}]} \funof{\freqp{\tftr[r]< \mathit{ss}_\mathit{th}}{f_b}{f_{\max}} \Rightarrow \freqp{\tftr[e]<e_{\max}}{0}{f_b}},
\end{equation}
where
\begin{itemize}
    \item $\mathit{ss}_\mathit{th}$ is the threshold value for the high-frequency components of the spectrum above which the signal is not considered to be in steady state, and
    \item $e_{\max}$ is the maximum acceptable steady-state error.
\end{itemize}

\subsubsection{Overshoot}
OV requirements constrain the extent to which an input frequency component can be amplified in the output.
Accordingly, we formulate them as an implication between a precondition capturing the presence of a frequency component in the input, and a postcondition that constrains its amplification in the output.
The precondition specifies a lower bound on the amplitude of an input
frequency component, while the postcondition defines an upper bound on
the allowable amplification of the corresponding output component.
Since the amplification usually concerns the spectrum around the bandwidth, we limit the predicate to a frequency interval around $f_b$.
The resulting \lang formula is:
\begin{equation}
    \square_{[0,t_{\max}]} \freqp{\tftr[r]>\mathit{ov}_{\mathit{th}} \Rightarrow \tftr[x]<\mathit{ov}_{\mathit{amp}}\cdot\tftr[r]}{f_b-\delta}{f_b+\delta},
\end{equation}
where
\begin{itemize}
    \item $\mathit{ov}_{\mathit{th}}$ is the threshold used to identify relevant frequency components in the input spectrum,
    \item $\mathit{ov}_{\mathit{amp}}$ is the maximum allowable amplification of the spectrum (e.g., $1.1$ corresponds to an OV of $10\%$), and
    \item $\delta$ defines the width of the frequency interval centred at the system bandwidth.
\end{itemize}

\subsubsection{Rise Time}
RT requirements are expressed in terms of the system bandwidth.
The system bandwidth defines a frequency threshold below which the output is expected to accurately track the input (i.e., same frequency components in the input and output) and above which input frequency components should be attenuated. 
Accordingly, we formulate RT requirements as the conjunction of two frequency-domain predicates, capturing the low- and high-frequency ranges.
For the low-frequency tracking, we use a predicate similar to that of the SS requirements.
However, because the system is not assumed to be at steady state, we allow a larger tracking error; thus a higher threshold for the allowable error.
For the high-frequency attenuation, we use a predicate similar to that of the OV requirements, consisting of a precondition that captures the presence of a frequency component in the input, and a postcondition that requires the corresponding output component to have a smaller amplitude.
Since frequency components around the system bandwidth may exhibit amplification, as discussed for the OV requirements, we exclude a frequency interval around the bandwidth from both predicates.
The resulting \lang formula is
\begin{equation}
    \square_{[0,t_{\max}]}
    \funof{\freqp{\tftr[e]<e_{\mathit{lf}}}{0}{f_b-\delta} \land
           \freqp{\tftr[r]>\mathit{rt}_{\mathit{th}} \Rightarrow \tftr[x]<\tftr[r]}{f_b+\delta}{f_{\max}}
          },
    \label{eq:rt-tf-template}
\end{equation}
where
\begin{itemize}
    \item $e_\mathit{lf}$ is the maximum allowable low-frequency tracking error at all times, and \item $\delta$ defines the width of the frequency interval around the system bandwidth that is excluded from the predicate. \end{itemize}

\subsubsection{Settling Time}
ST requirements can be formulated as a combination of bandwidth and overshoot constraints.
To avoid a repetition of the RT and OV requirements, we propose a slightly modified formulation of the ST requirements, that leverages a combination of time-domain and frequency-domain predicates.
The original requirement constrains the time the output takes to reach and remain within a specified tolerance of the input.
In contrast, our formulation constrains the time required for the output to reach steady state after a change in the input.
This can be naturally expressed using the ``until'' operator together with predicates that characterize whether the input and output are in steady state.
Specifically, whenever the input has remained at steady state for at least the settling time, the output is required to have also reached steady state, unless the input leaves the steady state before the settling time elapses.
As for the SS requirements, we formulate the steady-state condition by constraining the amplitude of high-frequency components.
The resulting \lang formula is
\begin{equation}
    \square_{[0,t_{\max}]}\funof{
        \freqp{\tftr[r]<ss_{th\_r}}{f_b}{f_{\max}} \mathcal{U}_{[0,\mathit{ST}]}
        \funof{\freqp{\tftr[x]<ss_{th\_x}}{f_b}{f_{\max}} \lor
               \freqp{\tftr[r]>ss_{th\_r}}{f_b}{f_{\max}}
            }
        },
\end{equation}
where
\begin{itemize}
    \item $\mathit{ST}$ is the prescribed settling time, and
    \item $\mathit{ss}_\mathit{th}$ is the threshold on the amplitude of high-frequency components above which the signal is not considered to be in steady state. This threshold is potentially different for the input and output signals, as the output can be affected by measurement noise.
\end{itemize}

     \section{Evaluation}
\label{sec:evaluation}
In this section, we compare the ability of the time and time-frequency domains to express signal-based requirements for CPSs.
First, we evaluate the {\em applicability} of the two domains, namely, their ability to formulate step-response specifications that are applicable to a larger set of traces.
As discussed in Section~\ref{sec:partitioning-req-classes}, step-response requirements are expressed as the combination of a precondition and a postcondition.
When the precondition is not satisfied, the specification yields a trivial verdict\textemdash either positive or negative, depending on how the specification is formulated.
Second, we evaluate the {\em fidelity} of the two domains, i.e., their ability to express the intended signal property.
For example, RT and ST requirements constrain how quickly the output tracks the input.
We therefore investigate whether the corresponding specifications are able to actually distinguish between systems with sufficiently fast responses and those with slower dynamics.
Third, we evaluate the {\em monitoring tolerance to signal noise} of the two domains, that is, the ability of the monitoring to provide consistent verdicts in presence of signal noise.
We express these three objectives with the following Research Questions (RQs):
\begin{enumerate}[label={\bf RQ\arabic*}]
    \item {\bf Broader Applicability}:
    What is the degree of applicability of the step-response requirements expressed in the time and time-frequency domains?
    \item {\bf Expression Fidelity}:
    What is the degree of fidelity to the intrinsic signal property of the requirements expressed in the time and time-frequency domains?
    \item {\bf Noise Tolerance of Monitoring}:
    What is the monitoring tolerance to signal noise of requirements expressed in the time and time-frequency domains?
\end{enumerate}

\subsection{Setup}
To answer our research questions, we need
\begin{enumerate*}[label=(\roman*)]
    \item test subjects to generate CPS input and output traces,
    \item input traces to exercise the test subjects, and
    \item specification languages, templates and monitors for expressing and checking the various classes of requirements.
\end{enumerate*}

\subsubsection{Test Subjects}
\label{sec:eval-test-subjects}
As test subjects, we consider two representative CPSs: a drone and a lightweight aircraft.
Since our specifications are defined over a single input-output pair, for the drone, we consider only the one-dimensional longitudinal dynamics rather than the full three-dimensional position.
Consequently, each test input specifies a desired back-and-forth movement of the drone, while the corresponding output is its actual longitudinal position.
To generate the traces, we use a simulation model of the Crazyflie drone, which has been shown in previous work to accurately reproduce the behaviour of the actual drone~\cite{mandrioli:2023a}.
For the lightweight aircraft, we use a publicly available simulation model developed by MathWorks, which has been also been adopted in previous software testing research~\cite{mandrioli:2023b}.\footnote{
    \url{https://nl.mathworks.com/help/aeroblks/lightweight-airplane-design.html}
}
The model includes an altitude controller.
Accordingly, each test input consists of a sequence of desired altitudes, and the corresponding output is the aircraft's actual altitude.

In addition to the drone and lightweight aircraft, answering RQ2 requires systems that intrinsically satisfy or violate the considered step-response requirements.
To obtain such systems, we rely on results from control theory, which provide analytical relationships between the parameters of linear system models and standard step-response characteristics.
Specifically, we use the formulas reported by \citet[Table~7.1, ch.~7, pg.~20]{Astrom:2008}, which explicitly relate between the model parameters to the values of SS, RT, OV, and ST.
Using these formulas, we can systematically generate systems that intrinsically either satisfy or violate a given requirement by construction.
Importantly, whether a system satisfies a requirement is {\em independent of the specific input} signal.
This allows us to generate multiple input-output traces from the same system, all of which consistently satisfy or violate the requirement.

\subsubsection{Generation of Input Traces}
\label{sec:eval-input-traces}
To generate input traces, we use two complementary approaches.
First, we generate traces that contain only a single step-like change in the input.
These traces are important because they match the input shape assumed in the definition of the step-response requirement classes.
Second, we generate traces that are representative of those produced by state-of-the-art CPS testing approaches.
Specifically, we randomly sample control points\textemdash i.e., time and value pairs\textemdash and then interpolate between them using different interpolation methods~\cite{Mandrioli:2025, Yadav:2025, Hildebrandt:2020, Matinnejad:2019}.
We consider three interpolation methods: linear interpolation (the signal changes linearly between two consecutive control points), constant interpolation (the signal value is held constant until next control point), and alternated interpolation (the signal changes linearly during first half of the interval between two control points and remains constant during the second half).

\subsubsection{Specification Languages and Templates}
\label{sec:setup-langs}
To compare the expression of CPS requirements in the time and time-frequency domains, we use the specification language STL*, proposed in~\citet{Brim:2014}, and \lang, introduced in this work.
STL* extends STL (presented in Section~\ref{sec:stl}) with a signal-value-freezing operator.
Simply put, this operator enables the comparison of values of the same signal at different points in time.
For the time-domain representation of signals, we choose STL* both because it is the standard language for specifying CPS properties, and because~\citet{Kapinski:2016b} provides templates for the requirement classes considered in this work.\footnote{
    Strictly speaking, the authors of~\citet{Kapinski:2016b} use plain STL.
    However, in the OS specification, they assume access to the value of a signal after a given number of time steps.
    To support this assumption while also obtaining more general versions of the formulas, we instead use STL*~\cite{Brim:2014}.
}
Below, we report the STL* formulas templates as used in our empirical evaluation. These are adapted from the STL templates proposed by Kapinski et al.~\cite{Kapinski:2016b}.
We remark that in the original work~\cite{Kapinski:2016b}, the formulas of the step-response requirements are written to be satisfied when a violation occurs.
To simplify the comparison with the formulas presented in Section~\ref{sec:templates}, here, we instead report equivalent formulas that are satisfied when the corresponding requirement is satisfied, documenting the changes that we applied.
In the formulas below, we use the same conventions as in Section~\ref{sec:templates}: $r$ is the input signal, $x$ is the output signal, and $t_{\max}$ is the total signal duration.

\partparagraph{SP}
To detect spikes, we use the same formula proposed by~\citet{Kapinski:2016b}
\begin{equation}
    \Diamond_{[0,t_{\max}]}((\dot{x}>m)\land \Diamond_{[0,w]}(\dot{x}<-m)),
\end{equation}
where
\begin{itemize}
    \item $\dot{x}$ is the discrete derivative of the signal computed as $(x-\mathit{prev}(x))/\mathit{dt}$ using the $\mathit{prev}$ operator that returns the value of a variable at the previous time step,
    \item $w$ is related to the spike width, and
    \item $m\cdot w$ is proportional to the spike amplitude.
\end{itemize}

\partparagraph{OS}
To detect oscillations, the formula proposed by~\citet{Kapinski:2016b} provides only an upper bound to the oscillation period.
We use instead a formula adaptation that allows both upper and lower bounds on the period
\begin{equation}
    \Diamond_{[0,t_{\max}]}(\mathit{up}\land \Diamond_{[2d-\varepsilon,2d+\varepsilon]}\mathit{down}),
\end{equation}
where
\begin{itemize}
    \item $\mathit{up}$ and $\mathit{down}$ are respectively defined as $\ast_d x-x>a$ and $x-\ast_d x>a$, using the signal-value-freezing operator $\ast_d$ from STL* to access the future value $x(t+d)$ at time $t$
    \item $a$ defines the minimum amplitude of the oscillation,
    \item $4d$ corresponds to the oscillation period, and
    \item $\varepsilon$ is a small number used to have some margin in the assessment of the period.
\end{itemize}
Compared to the original formula, we changed the time interval of the $\Diamond_{[2d-\varepsilon,2d+\varepsilon]}$, which in the original paper is $\Diamond_{[0,2d]}$.
This introduces a tighter bound in the distance between the $\mathit{up}$ and $\mathit{down}$ predicates, thus allowing a more precise specification of the period.

\partparagraph{SS}
To check the SS requirement, we use the formula proposed by~\citet{Kapinski:2016b} and simply negate it
\begin{equation}
    \neg\Diamond_{[0,t_{\max}]}\left((\abs{\dot{r}}<\varepsilon\land \abs{\dot{x}}<\varepsilon) \mathcal{U}_{[\tau,\tau+\delta]} e>e_{\max}\right),
    \label{eq:ss-stl}
\end{equation}
where
\begin{itemize}
    \item $\dot{r}$ and $\dot{x}$ are the input and output derivatives, defined as above,
    \item $\varepsilon$ is the derivative threshold below which the input and output are considered constant,
    \item $\tau$ is the expected settling time, i.e., the time after
      which the output is expected to have reached steady state. The interval $[\tau,\tau+\delta]$ defines the observation window over which the steady-state error is evaluated. 
     \item $e_{\max}$ is the maximum allowable steady-state error.
\end{itemize}

\partparagraph{OV}
To check the OV requirements, we use the formula proposed by~\citet{Kapinski:2016b} and negate the constraint after the step detection.
The adapted formula is
\begin{equation}
    \Diamond_{[0,t_{\max}]}(\mathit{step}(r)\land \square_{[0,\tau]}e<\mathit{ov}_{\max}),
\end{equation}
where
\begin{itemize}
    \item $\mathit{step}(r)=\abs{r-\mathit{prev}(r)}>\sigma$ checks whether a step-like change of amplitude at least $\sigma$ has occurred in the input signal $r$,
    \item $\tau$ is the expected settling time. 
    \item $\mathit{ov}_{\max}$ is the maximum allowable overshoot, usually computed proportionally to the step size, e.g., $\mathit{ov}_{\max}=0.2\mathit{step\_size}(r)$ corresponds to allowing a $20\%$ overshoot with respect to the step size $\mathit{step\_size}(r)=r-\mathit{prev}(r)$.
\end{itemize}
Compared to the original work~\cite{Kapinski:2016b}, we changed $\square_{[0,\tau]}e<\mathit{ov}_{\max}$, that negates the original $\Diamond_{[0,\tau]}e>\mathit{ov}_{\max}$.

\partparagraph{RT}
To check the RT requirements, we use the formula proposed
by~\citet{Kapinski:2016b} and negate the constraint after the step detection.
The adapted formula is
\begin{equation}
    \Diamond_{[0,t_{\max}]}(\mathit{step}(r)\land \Diamond_{[0,\mathit{RT}]}x>0.75\cdot\mathit{step\_size}(r)) ,
    \label{eq:rt-time}
\end{equation}
where
\begin{itemize}
    \item $\mathit{RT}$ is the required rise time, and
    \item $0.75\mathit{step\_size}(r)$, using the same definition of $\mathit{step\_size}(r)$ as above, is the threshold at which the rise time is checked.\footnote{
        This definition measures the rise time from $0\%$ to $75\%$ of the step size.
        As described in Section~\ref{sec:background-requirements}, the standard definition measures the rise time between $10\%$ and $90\%$ of the step size.
        Intuitively, the two quantities are closely related and can be converted approximately using the relation $RT_{0-75\%}\approx 0.65RT_{10-90\%}$.
    }
\end{itemize}
As remarked above, compared to the original work~\cite{Kapinski:2016b}, we changed $\Diamond_{[0,\mathit{RT}]}x>0.75\cdot\mathit{step\_size}(r))$, that negates the original $\square_{[0,\mathit{RT}]}x<0.75\cdot\mathit{step\_size}(r))$.
Further, we note that Eq.~\ref{eq:rt-time} implicitly assumes that the baseline output value before the step is zero (or has been de-trended to zero), we discuss the implications of this implicit assumption for our results in the threats to validity~\ref{sec:threats-internal}.

\partparagraph{ST}
To check the ST requirements, we use the formula proposed
by~\citet{Kapinski:2016b} and negate the constraint after the step detection.
The adapted formula is
\begin{equation}
    \Diamond_{[0,t_{\max}]}(\mathit{step}(r)\land \square_{[\mathit{ST},2\mathit{ST}]} (e<\beta\mathit{step\_size}(r))),
\end{equation}
where
\begin{itemize}
    \item $\beta$ specifies the allowable error band once the settling time has elapsed, expressed as a fraction of the input step size, and 
    \item $\mathit{ST}$ is the required settling time.
\end{itemize}
Compared to the original work~\cite{Kapinski:2016b}, the modified part is $\square_{[\mathit{ST},2\mathit{ST}]} (e<\beta\mathit{step\_size}(r))$, that negates the original $\Diamond_{[\mathit{ST},t_{\max}]} (e>\beta\cdot\mathit{step\_size}(r))$.
Besides the negation, we also changed the time interval of the temporal operator from $[ST,t_{\max}]$ to $[\mathit{ST},2\mathit{ST}]$.
This makes the adapted formula more general, as it does not need to assume that the signal remains unchanged after the step.

For the time-frequency domain part of the experiments, we choose \lang, proposed in this work, together with the formula templates presented in Section~\ref{sec:templates}. Using \lang requires selecting a time-frequency transform. In our experimental evaluation, as discussed in Section~\ref{sec:background-time-freq}, we use the STFT because it is simple and easy-to-interpret. When using the STFT, both the window function and window length must be specified. Since these parameters directly affect the semantics of the formulas, they must be chosen according to the application. Accordingly, the methodology for each RQ discusses the selected window shapes and lengths.

\subsubsection{Monitoring Tool}
To monitor the traces, we use the extended version of
RTAMT~\cite{Nickovic:2020,yamaguchi2024} presented in
Section~\ref{sec:lang-implementation}, which supports the monitoring of both STL* and \lang.
In this way, we use the same monitoring framework for both STL* and \lang formulas, ensuring a fair comparison.

\subsubsection{Settings}
We ran the experiments on a MacBook Pro with M1 chip and \qty{32}{\giga\byte} of memory running macOS 26 and Python 3.12.13.
The experiments took around \qty{14.5}{\hour}.

\subsection{RQ1: Broader Applicability}
\label{sec:rq1}
\subsubsection{Methodology and Parameters Settings}
To answer this RQ, we assess how many CPS test-input traces satisfy the preconditions of the step-response requirements.
To this end, we generated test-input traces for our drone and lightweight aircraft test subjects using the input shapes discussed in Section~\ref{sec:eval-input-traces}.
We then evaluated whether these traces satisfy the corresponding specification preconditions.
For the time-domain specifications, the precondition is identical for the ST, OV, and RT requirements: the input must exhibit a step-like change.
We verified this by monitoring the formula
\begin{equation}
    \Diamond_{[0,t_{\max}]} \abs{r-\mathit{prev}(r)}>\sigma
    \label{eq:stl-step-precondition}
\end{equation}
over the input signal $r$. This formula corresponds to the definition of $\mathit{step}(r)$ introduced in Section~\ref{sec:setup-langs}.
The SS requirement has a different precondition: the input trace must satisfy the input component of the first operand of the until operator.
We verified this by monitoring the formula
\begin{equation}
    \Diamond_{[0,t_{\max}]} \square_{[0,\tau]} \abs{\dot{r}}<\varepsilon.
    \label{eq:stl-ss-precondition} 
\end{equation}

For the time-frequency-domain specifications, the precondition differs across requirement classes.
For SS, we checked that the input is at steady state by verifying that the amplitude of the high-frequency components remains below a threshold:
\begin{equation}
    \Diamond_{[0,t_{\max}]} \freqp{\tftr[r]< \mathit{ss}_\mathit{th}}{f_b}{f_{\max}} .
\end{equation}
For OV, we checked that at least one input frequency component around the bandwidth is sufficiently large to trigger a transient response in the CPS, thereby enabling the assessment of overshoot:
\begin{equation}
    \Diamond_{[0,t_{\max}]} \neg \freqp{\tftr[r]<\mathit{ov}_{\mathit{th}}}{f_b-\delta}{f_b+\delta} .
\end{equation}
Similarly, for RT, we checked that at least one high-frequency component above the bandwidth is sufficiently large to enable the assessment of high-frequency filtering:
\begin{equation}
    \Diamond_{[0,t_{\max}]} \neg \freqp{\tftr[r]<\mathit{rt}_{\mathit{th}}}{f_b+\delta}{f_{\max}} .
\end{equation}
Finally, for ST, we checked that the input remains at steady-state for at least the settling time:
\begin{equation}
    \Diamond_{[0,t_{\max}]} \square_{[0,\mathit{ST}]} \freqp{\tftr[r]<\mathit{ss}_{\mathit{th\_r}}}{f_b+\delta}{f_{\max}} .
\end{equation}
Appendix~\ref{sec:app-rq1-params} reports the parameters values used in the precondition formulas and explains how we selected each value for our test subjects.

For the time-frequency transform, we used a rectangular window of length $\qty{15}{\second}$ for the lightweight aircraft, and $\qty{4}{\second}$ for the drone.
Intuitively, the aircraft exhibits slower dynamics, and therefore requires a longer time window.
More generally, the window should be long enough to capture the system dynamics, for example, the time required for the drone to change direction after receiving a new target position.

We generated $200$ input traces for each input type, yielding a total of $800$ input traces per test subject.
We then monitored each precondition specification over all traces and counted how many satisfy the corresponding precondition when the requirement is expressed in STL and in \lang, respectively.

\begin{table}
    \caption{RQ1: Satisfaction of the STL* and \lang Precondition in the Test Inputs of the Different Test Subjects.
    For each input type (corresponding to the columns), we generated 200 input traces.
    The reported percentage indicates how many of those traces satisfy
    each precondition (corresponding to the rows).}
    \label{tab:rq1-applicability}
    \begin{tabular}{llccccc}

\toprule
\multirow{2}{*}{\textbf{Test Subject}}
& \multirow{2}{*}{\textbf{Precondition}}
& \multicolumn{4}{c}{\textbf{Input Type}}
& \multirow{2}{*}{\textbf{Total}}
\\

  \cline{3-6}
&
& \textbf{Step} 
& \textbf{Rand-Const}
& \textbf{Rand-Linear}
& \textbf{Rand-Alt}
& \\
\midrule

\multirow{5}{*}{\makecell{Drone\\Crazyflie}}
& STL* Step & $100.0\%$ & $ 97.0\%$ & $  1.0\%$ & $  0.0\%$ & $ 49.5\%$ \\
& STL* SS   & $100.0\%$ & $100.0\%$ & $100.0\%$ & $100.0\%$ & $100.0\%$ \\
& \lang RT  & $100.0\%$ & $ 99.0\%$ & $100.0\%$ & $ 99.5\%$ & $ 99.6\%$ \\
& \lang ST  & $100.0\%$ & $100.0\%$ & $100.0\%$ & $100.0\%$ & $100.0\%$ \\
& \lang OV  & $100.0\%$ & $ 99.5\%$ & $100.0\%$ & $100.0\%$ & $ 99.9\%$ \\
& \lang SS  & $100.0\%$ & $100.0\%$ & $100.0\%$ & $100.0\%$ & $100.0\%$ \\

\midrule

\multirow{5}{*}{\makecell{Lightweight\\Aircraft}} 
& STL* Step   & $ 89.0\%$ & $100.0\%$ & $ 10.0\%$ & $ 17.0\%$ & $ 54.0\%$ \\
& STL* SS     & $100.0\%$ & $100.0\%$ & $100.0\%$ & $100.0\%$ & $100.0\%$ \\
& \lang RT    & $100.0\%$ & $100.0\%$ & $100.0\%$ & $100.0\%$ & $100.0\%$ \\
& \lang ST    & $100.0\%$ & $100.0\%$ & $100.0\%$ & $100.0\%$ & $100.0\%$ \\
& \lang OV    & $100.0\%$ & $100.0\%$ & $100.0\%$ & $100.0\%$ & $100.0\%$ \\
& \lang SS    & $100.0\%$ & $100.0\%$ & $100.0\%$ & $100.0\%$ & $100.0\%$ \\

\bottomrule

\end{tabular} \end{table}

\subsubsection{Results}
Table~\ref{tab:rq1-applicability} reports the results of evaluating the CPS requirements preconditions on the generated input traces.
For each test subject, the rows report the results for the different preconditions, while the columns report the percentage of the $200$ traces with a given input type that satisfy the corresponding precondition.
For the preconditions expressed in \lang, the satisfaction values are always either $100\%$ or above $99\%$, demonstrating that these requirements are applicable to all the generated input traces.
Similarly, the STL* precondition for SS is satisfied by all traces.
In contrast, the STL* Step precondition is satisfied for almost all step and rand-const inputs\textemdash that is, traces generated using constant interpolation.
However, only a small fraction of the linearly interpolated (rand-linear) and alternated interpolation (rand-alt) traces satisfy this precondition, with satisfaction rates of $1\%$ and $0\%$ for the drone, and $10\%$ and $17\%$ for the lightweight aircraft, respectively. 
This result is expected because these input types change gradually and therefore rarely exhibit the abrupt (step-like) changes required by the STL* step precondition (Eq.~\ref{eq:stl-step-precondition}).

This comparison is particularly relevant because linearly interpolated traces are among the most commonly used inputs for CPS testing.
In fact, the abrupt step-like changes introduced by constant interpolation (and by step inputs) can place significant stress on CPS components, specifically the actuators, as they require sudden changes in the system state.
These results therefore highlight an important advantage of the time-frequency signal representation: it extends the specification of step-response requirements to a much broader class of input traces.

\begin{summarybox}
    {\bf Summary:} Our experiments show that expressing the ST, RT and OV requirements in \lang extends their applicability to input traces generated using linear or alternated interpolation between control points, compared to their STL* counterparts.
    Specifically, we observe that, for our test subjects, while only $0\%\text{--}17\%$ of the linear and alternated interpolation traces satisfy the corresponding time-based STL* preconditions, nearly all of them ($\geq99\%$) satisfy the time-frequency-based \lang preconditions.
    This applicability extension is particularly important because linear interpolation is the most commonly used method for generating CPS test inputs.
     At the same time, the time-frequency representation preserves applicability to step and constant-interpolation inputs.
\end{summarybox}

\subsection{RQ2: Requirements Expression Fidelity}
\label{sec:rq2}
\subsubsection{Methodology and Parameters Settings}
\label{sec:rq2-methodology}
To answer RQ2, we compare the ability of STL* and \lang specifications to detect requirement violations in signals that intrinsically satisfy or violate them.
In this way, we assess the specifications' fidelity to the intrinsic properties of the signals.
We compare the specifications in terms of \emph{precision} (the
ability to report only violations), and \emph{soundness} (the ability
to report all violations)~\cite{meyer2019}.

To build signals that intrinsically satisfy or violate the requirements, we adopted different approaches for the signal-feature and step-response requirement classes.
For the signal-feature requirements (SP and OS), we generated signals that either contain or do not contain the relevant feature (i.e., spikes and oscillations) with prescribed parameters.
To obtain a balanced dataset, we also generated an equal number of traces that either lack the feature altogether or contain it with parameters different from the prescribed ones (e.g., oscillations that are too slow or too fast).
We generated these violating signals based on common violation patterns identified in the literature on diagnosis of signal-based requirements~\cite{Chaima:2023}.
For SP, we considered three spike shapes: rectangular, triangular, and Gaussian.
For each spike shape, we generated traces of length \qty{20}{\second} containing a single spike with prescribed ranges of amplitude values (between $10$ and $20$) and duration (between $\qty{1}{\second}$ and $\qty{2}{\second}$).
Following~\citet{Chaima:2023}, we also generated invalid spikes that are either too small (amplitude between $0$ and $10$) or too wide (duration between $\qty{2}{\second}$ and $\qty{4}{\second}$).
Traces that do not contain spikes instead consist of increasing, decreasing, or constant signals.
For OS, we considered three oscillation shapes: sinusoidal, rectangular, and triangular.
For each oscillation shape, we generated traces of length \qty{20}{\second} containing an oscillatory segment lasting between \qty{5}{\second} and \qty{10}{\second}, while the remainder of the trace is constant.
The oscillations have a prescribed frequency between \qty{0.9}{\hertz} and \qty{1.1}{\hertz} and an amplitude between $10$ and $20$.
We generated invalid oscillations that are either too large (amplitude between $20$ and $40$), too small (amplitude between $5$ and $10$), too slow (frequency \qty{0.09}{\hertz} and \qty{0.9}{\hertz}), or too fast (frequency between \qty{1.1}{\hertz} and \qty{11}{\hertz}).
As for SP, following~\citet{Chaima:2023}, we also generated traces that do not contain oscillations; these consist of increasing, decreasing, or constant signals.
For each violation type and feature shape, we generated $30$ traces, resulting in a total of $540$ traces for SP and $900$ for OS.
As mentioned above, half of the traces contain a valid feature (equally distributed across the three shapes), while the other half either contain an invalid feature or no feature at all (equally distributed across the different shapes and violation types).
Since the OS requirement class includes more types of invalid traces than SP, it requires a larger dataset; thus we generated more traces for this class.

\begin{table}
    \centering
    \caption{RQ2: Threshold values used to instantiate the specifications for each step-response requirement class, together with the corresponding response feature values used to synthesise satisfying and violating systems.}
    \label{tab:rq2-sys-values}
    \footnotesize
\begin{tabular}{lccc}
\toprule
  \textbf{Class (Response feature)}
& Specification Threshold
& Satisfying System
& Violating System
\\
\midrule

RT (Rise Time)
& $\qty{0.75}{\second}$
& $\qty{0.5}{\second}$
& $\qty{1}{\second}$
\\

ST (Settling Time)
& $\qty{4.5}{\second}$
& $\qty{3}{\second}$
& $\qty{6}{\second}$
\\

OV (Overshoot)
& $15\%$
& $10\%$
& $20\%$
\\

SS (Steady-State Error)
& $0.1$
& $0.05$
& $0.2$
\\

\bottomrule

\end{tabular}
 \end{table}

For the step-response requirements, we instead require traces containing both a test input and the corresponding output.
Therefore, as described in Section~\ref{sec:eval-test-subjects}, we used control theory to generate synthetic system models that either satisfy or violate the requirements.
We then simulated these systems using different input signals to obtain satisfying and violating input-output traces.
For each requirement class, we first selected a threshold value for the corresponding response feature (e.g., a desired rise time of \qty{0.75}{\second} for the RT class) that we used to instantiate the corresponding specification formula.
We then synthesized two system models: one whose response exhibits a feature value that is strictly better than the threshold, and therefore satisfies the specification (e.g., a rise time of \qty{0.5}{\second}); and another whose response exhibits a feature value that is worse than the threshold, and therefore violates the specification (e.g., a rise time of \qty{1}{\second}).
By simulating each system under different input signals, we obtain input-output traces that intrinsically satisfy or violate the corresponding requirement.
Table~\ref{tab:rq2-sys-values} summarizes, for each requirement class, the selected threshold together with the response feature values exhibited by the synthesized satisfying and violating systems.

Since, as mentioned above, assessing specifications on traces to which they are not applicable would lead to a trivial verdict, we considered only input traces to which both STL* and \lang specifications are applicable.
Based on the results of RQ1, for ST, RT, and OV requirement classes, we considered only step and constant-interpolation inputs.
For SS, instead, we also considered linear and alternated interpolations.
For each input type, we generated $30$ traces, resulting in a total of $60$ input traces for each system used to evaluate ST, RT, and OV; and $120$ input traces for each system used to evaluate SS. For all the traces, we set a duration of \qty{100}{\second}, and randomly generated amplitude values (for the step-like changes and input control points) up to $10$.

\begin{table}
    \centering
    \caption{RQ2: STL* and \lang formulas used for RQ2, including the parameters values.}
    \label{tab:rq2-specs}
    \tiny
    \begin{tabular}{ccl}
\toprule
  {\bf Req.}
& {\bf Lang.}
& {\bf Formula}
\\
\midrule

 \multirow{2}{*}{SP}
& STL*
& $ \Diamond_{[0,20]} ((x-\mathit{prev}(x))/0.01>5 \land \Diamond_{[0,2]} (x-\mathit{prev}(x))/0.01<-5)$
\\
& \lang
& $ \Diamond_{[0,20]} \square_{[0,1]} \neg \freqp{\tftr[x]<1}{0.5}{4} $
\\ \cline{2-3}

 \multirow{2}{*}{OS}
& STL*
& $ \Diamond_{[0,20]} ((\ast_{0.25}x-x)>20 \land (\ast_{0.25}x-x)<40 \land (\Diamond_{[0.48,0.52]} ((x-\ast_{0.25}x)>20 \land (x-\ast_{0.25}x)<40)))$
\\
& \lang
& $ \Diamond_{[0,20]} (\neg \freqp{\tftr[x]<4.5 \lor \tftr[x]>10}{0.85}{1.15}\land \freqp{\tftr[x]<1}{0.4}{0.85}\land \freqp{\tftr[x]<1}{1.15}{1.6})$
\\ \cline{2-3}

 \multirow{2}{*}{RT}
& STL*
& $ \Diamond_{[0,100]} (\abs{r-\mathit{prev}(r)}>1 \land (\Diamond_{[0,0.75]} x<0.75r))$
\\
& \lang
& $ \square_{[0,100]} (\freqp{\tftr[r-x]<0.25}{0}{0.3} \land \freqp{\tftr[r]>0.1 \Rightarrow \tftr[x]<\tftr[r]}{0.6}{4})$
\\ \cline{2-3}

 \multirow{2}{*}{ST}
& STL*
& $ \Diamond_{[0,100]} (\abs{r-\mathit{prev}(r)}>1 \land \square_{[4.5,6]} e<0.02r) $
\\
& \lang
& $ \square_{[0,100]} (\freqp{\tftr[r]<0.1}{0.15}{4} \mathcal{U}_{[0,2.25]} (\freqp{\tftr[x]<0.1}{0.15}{4} \lor \neg\freqp{\tftr[r]<0.1}{0.15}{4}))$
\\ \cline{2-3}

 \multirow{2}{*}{OV}
& STL*
& $ \Diamond_{[0,100]} (\abs{r-\mathit{prev}(r)}>1 \land (\square_{[0,2]} (x-r)<0.15r)) $
\\
& \lang
& $ \square_{[0,100]} \freqp{\tftr[x]>0.05 \Rightarrow \tftr[x]<1.15\tftr[r]}{1}{3} $
\\ \cline{2-3}

 \multirow{2}{*}{SS}
& STL*
& $ \neg\Diamond_{[0,100]} (((r-\mathit{prev}(r))/0.01<0.01 \land (x-\mathit{prev}(x))/0.01<0.01) \mathcal{U}_{[2,2.02]} e>0.1) $
\\
& \lang
& $ \square_{[0,100]} (\freqp{\tftr[r]<0.01}{0.5}{4} \Rightarrow \freqp{\tftr[e]<0.1}{0}{0.2})$
\\

\bottomrule

\end{tabular} \end{table}

In Table~\ref{tab:rq2-specs}, we report the complete specification formulas used for this RQ.
These formulas are based on the templates discussed in Section~\ref{sec:templates} for \lang and Section~\ref{sec:setup-langs} for STL* (i.e.,
on~\citet{Kapinski:2016b}).
We chose the threshold values based on the value ranges of the valid signals discussed above or on known system parameters (e.g., the expected bandwidth $f_b$).
In Appendix~\ref{sec:app-rq2-params}, we detail how we selected each parameter of every formula for both STL* and \lang.
Here, we highlight one important rule of thumb that guided our parameter selection.
When defining the thresholds for frequency-domain values, low-frequency components usually have amplitudes comparable to those of the corresponding time-domain signal, while higher-frequency components are usually one to two orders of magnitude smaller. 
Intuitively, fast changes in CPS traces tend to be less persistent than slow changes, resulting in lower amplitudes in the frequency domain.

For monitoring the \lang formulas, we used different time-frequency transforms depending on the requirement.
In general, the window length must be sufficient to both capture the relevant system dynamics and provide adequate frequency resolution\textemdash as mentioned in Section~\ref{sec:background-time-freq}, a longer time window yields higher frequency resolution.
For the signal-feature requirements (SP and OS), which only need to detect the presence of a signal feature over $\qty{20}{\second}$ traces, we used a triangular window of length $\qty{5}{\second}$.
For ST and SS, we used a rectangular window because these requirements
define assertions on slower signal dynamics, such as when the signals are at steady state (i.e., approximately constant signals).
For OV and RT, we used a triangular window as they predicate on events that occur during the transient and therefore require frequency information about the signals more localised in time.
For the step-response requirements, the system dynamics are on the
order of a few seconds (corresponding to frequencies in the order of
\unit{\hertz}); therefore, to achieve sufficient frequency resolution,
we used a window of length $\qty{24}{\second}$.

After generating the traces and monitoring them, we compute the precision and soundness of both STL* and \lang formulas.
We compute the precision as the percentage of traces detected as violating that are actual violations, i.e., $\text{Precision}=c/(c+f)$, where $c$ is the number of correctly caught violations and $f$ is the number of non-violating traces incorrectly flagged as violating.
We compute the soundness as the percentage of all violating traces that are detected, i.e., $\text{Soundness}=c/(c+m)$, where $c$ is the same as for the precision and $m$ is the number of missed violations.

\begin{table}
    \centering
    \caption{RQ2: Requirements Expression Fidelity of STL* and \lang for Different Requirement Classes.}
    \label{tab:rq2-fidelity}
    \begin{tabular}{ccccc}

\toprule
\multirow{2}{*}{\makecell[c]{\bf Requirement\\\bf Class}}
& \multicolumn{2}{c}{\bf STL*}
& \multicolumn{2}{c}{\bf \lang}
\\

  \cline{2-3}
  \cline{4-5}
& \textbf{Precision}
& \textbf{Soundness}
& \textbf{Precision}
& \textbf{Soundness}
\\
\midrule

SP    &    $67.84\%$ &  $100.00\%$ &    $81.48\%$ & $97.78\%$ \\\midrule
OS    &    $71.57\%$ &   $48.67\%$ &    $80.11\%$ & $94.00\%$ \\\midrule
RT    &    $92.68\%$ &   $63.33\%$ &    $76.60\%$ & $60.00\%$ \\\midrule
ST    &    $93.33\%$ &   $93.33\%$ &   $100.00\%$ & $53.33\%$ \\\midrule
OV    &    $93.33\%$ &   $70.00\%$ &    $87.50\%$ & $81.67\%$ \\\midrule
SS    &    $58.17\%$ &   $74.17\%$ &    $89.66\%$ & $65.00\%$ \\\midrule
Total &    $71.19\%$ &   $70.00\%$ &    $82.12\%$ & $86.47\%$ \\  

\bottomrule

\end{tabular}
 \end{table}

\subsubsection{Results}
\label{sec:rq2-results}
Table~\ref{tab:rq2-fidelity} reports the precision and soundness of the STL* and \lang specifications when detecting spikes and oscillations (SP and OS rows, respectively), and assessing the step-response requirements (RT, ST, OV, and SS).
The last row (``Total'') summarizes the results across all traces.

Overall, the last row shows that the \lang specifications achieve higher precision and soundness than the STL* specifications.
Specifically, \lang specifications attain a precision of $82.12\%$ and a soundness of $86.47\%$, compared with $71.19\%$ and $70\%$, respectively, for the STL* specifications.
However, the comparison across the individual requirement classes is more nuanced.

For SP, both STL* and \lang achieve very high soundness ($97.78\%$ for \lang and $100\%$ for STL*). 
However, \lang attains higher precision ($81.48\%$ compared with $67.84\%$ for STL*), indicating that STL* detects more spikes than it should.
This is due to its reliance on the derivative, which limits its
ability to define assertions on the actual amplitude of the spike and, for example, leads to the incorrect detection of too small spikes.
For OS, \lang clearly outperforms STL* in both soundness ($94\%$ for \lang against $48.67\%$ for STL*) and precision ($80.11\%$ for \lang against $71.57\%$ for STL*).
This result is expected, as the frequency domain is well suited for describing oscillatory behaviour.

For the step-response requirement classes, the comparison is less unilateral.
For RT, STL* specifications outperform the \lang specifications, particularly in terms of precision ($76.60\%$ for \lang against $92.78\%$ for STL*), while the soundness is comparable ($60\%$ for \lang against $63.33\%$ for STL*).
For the remaining classes, the two languages exhibit different trade-offs between soundness and precision.
For ST, \lang achieves higher precision ($100\%$ against $93.33\%$ for STL*), but substantially lower soundness ($53.33\%$ against $93.33\%$ for STL*).
Similarly, for SS, \lang provides significantly higher precision ($89.66\%$ against $58.17\%$ for STL*), but lower soundness ($65\%$ against $74.17\%$ for STL*).
In contrast, for OV, the trend is reversed: \lang achieves higher soundness ($81.67\%$ against $70\%$ for STL*), while STL* attains higher precision ($93.33\%$ against $87.50\%$ for \lang).

\begin{table}
    \centering
    \caption{RQ2: Requirements Expression Fidelity of \lang for ST, RT, and OV requirement classes, using linear and alternate interpolation inputs.}
    \label{tab:rq2-fidelity-extra}
    \begin{tabular}{lcc}

\toprule
\multirow{2}{*}{\makecell[l]{\bf Requirement Class}}
& \multicolumn{2}{c}{\bf \lang}
\\

  \cline{2-3}
& \textbf{Precision}
& \textbf{Soundness}
\\
\midrule

RT &     $92.86\%$ & $65.00\%$ \\\midrule
ST &    $100.00\%$ & $66.67\%$ \\\midrule
OV &    $100.00\%$ & $ 5.00\%$ \\

\bottomrule

\end{tabular}
 \end{table}

Overall, for the step-response requirement classes, neither specification language clearly outperforms the other.
However, for RT, ST and OV, it is important to take into account the
results of RQ1, which showed that the \lang specifications are
applicable also to test cases (i.e. input traces) generated using
linear and alternated interpolation.
In practice, this means that the STL* specifications can be more closely tailored to the specific traces and can therefore make stronger assumptions about the signal shapes.
In Table~\ref{tab:rq2-fidelity-extra}, we report the precision and soundness of the \lang specifications on test cases with linear and alternated interpolation.
Across all three requirement classes, we observe high precision but comparatively lower soundness.
Notably, the soundness for OV is very low, with a detection of only $5\%$ of the violations.
This is due to the nature of the overshoot requirement, which requires a strong excitation of the system to expose a violation.
This excitation is more readily produced by step-like changes in the input rather than by linear or alternated interpolation.

\begin{summarybox}
    {\bf Summary:} Our results show that, for detecting spikes and oscillations (the signal-feature requirement classes), \lang outperforms STL* in both soundness and precision.
    In contrast, for the step-response requirement classes, the two specification languages exhibit comparable overall performance, although they generally present different trade-offs between soundness and precision.
    For RT, STL* outperforms \lang, achieving both higher precision and soundness.
    For ST, STL* maintains a balanced trade-off between precision and soundness, whereas \lang achieves higher precision compared to soundness.
    For OV, the opposite trend is observed: STL* favors precision over soundness, while \lang provides a more balanced trade-off between the two. 
    Finally, for SS, STL* achieves higher soundness, whereas \lang attains higher precision.
\end{summarybox}

\subsection{RQ3: Noise Tolerance of Monitoring}
\label{sec:rq3}
\subsubsection{Methodology}
\label{sec:rq3-methodology}
To answer this RQ, we inject increasing levels of noise into the traces and measure for which noise level the verdict changes.
To compare the noise tolerance of the STL* and \lang specifications, we selected from the traces used in RQ2 only those for which the specifications in both languages provide the correct verdict with respect to the ground truth, where the ground truth is given by whether each trace intrinsically satisfies or violates the requirement. 
Intuitively, if the initial verdict for the non-noisy trace is incorrect, a change is not necessarily undesirable, even if due to noise.
In fact, the verdict would change to the correct one, and the noise could be interpreted to be ``helping'' the monitoring process rather than impairing it.
However, whether this is desirable or not depends on the application context and therefore we excluded this scenario from the analysis.
After applying this filtering step, we retained $379$ traces for SP, $538$ traces for OS, $65$ traces for RT, $87$ traces for ST, $82$ traces for OV, and $110$ traces for SS.

We injected three standard types of noise commonly observed in CPS signals, typically arising from sensor noise or calibration errors: constant offset, Gaussian white noise, and Gaussian high-frequency noise.
We define the noise level as a percentage of the signal's value range.
Specifically, for each signal, we computed its value range (i.e., the difference between its maximum and minimum values) and used a percentage of that range to define either the offset magnitude or the standard deviation of the Gaussian noise.
We considered four equally spaced noise levels ranging from $5\%$ to $30\%$.

For the step-response specifications, we injected noise only into the output signal, as the input signal is generated by a user or other software components and is therefore not subject to sensor noise.
However, this means that, for the SS requirement class, the offset affects the intrinsic satisfaction of the requirement or, in other words, affects the ground truth, thus making a verdict change actually desirable.
A similar argument applies to the \lang formulation of RT requirements, which constrain the low-frequency correspondence between the input and output (Eq.~\ref{eq:rt-tf-template}).
Since, in these cases, the information in the traces that is relevant to the requirement satisfaction changes, a verdict change may be actually desirable.
For this reason, we excluded these scenarios from the analysis of noise tolerance of monitoring.

\begin{table}
    \centering
    \caption{RQ3: Noise Tolerance of Monitoring.
    For each requirement class, we report the percentage of traces whose verdict changes due to noise (\% Flipped) and the average noise level at which the verdict change occurs (Avg. Flip Level ).}
    \label{tab:rq3-noise-tolerance}
    \begin{tabular}{cc cc cc cc}

\toprule
\multirow{2}{*}{\makecell[c]{\bf Requirement\\\bf Class}}
& \multirow{2}{*}{\bf Metric}
& \multicolumn{2}{c}{\bf Offset}
& \multicolumn{2}{c}{\bf White Noise}
& \multicolumn{2}{c}{\bf High-Freq. Noise}
\\

  \cline{3-8}
&
& \textbf{STL*}
& \textbf{\lang}
& \textbf{STL*}
& \textbf{\lang}
& \textbf{STL*}
& \textbf{\lang}
\\
\midrule

\multirow{2}{*}{SP} &  $\%$ Flipped    & $    0.00\%$ & $    2.90\%$ & $   30.08\%$ & $    3.69\%$ & $   30.08\%$ & $    0.00\%$ \\ \cline{2-8}
                    &  Avg. Flip Level & $          $ & $     0.202$ & $     0.058$ & $     0.199$ & $     0.057$ & $          $ \\ \cline{1-8}
\multirow{2}{*}{OS} &  $\%$ Flipped    & $    0.00\%$ & $    0.37\%$ & $   26.58\%$ & $   31.04\%$ & $   24.72\%$ & $    0.00\%$ \\ \cline{2-8}
                    &  Avg. Flip Level & $          $ & $     0.300$ & $     0.228$ & $     0.232$ & $     0.234$ & $          $ \\ \cline{1-8}
\multirow{2}{*}{RT} &  $\%$ Flipped    &         N/A  &         N/A  & $   29.23\%$ & $   67.69\%$ & $   29.23\%$ & $    6.15\%$ \\ \cline{2-8}
                    &  Avg. Flip Level &         N/A  &         N/A  & $     0.125$ & $     0.133$ & $     0.116$ & $     0.279$ \\ \cline{1-8}
\multirow{2}{*}{ST} &  $\%$ Flipped    & $   64.37\%$ & $    0.00\%$ & $   64.37\%$ & $   57.47\%$ & $   64.37\%$ & $    0.00\%$ \\ \cline{2-8}
                    &  Avg. Flip Level & $     0.050$ & $          $ & $     0.050$ & $     0.205$ & $     0.050$ & $          $ \\ \cline{1-8}
\multirow{2}{*}{OV} &  $\%$ Flipped    & $   60.98\%$ & $   10.98\%$ & $   60.98\%$ & $   59.76\%$ & $   60.98\%$ & $    2.44\%$ \\ \cline{2-8}
                    &  Avg. Flip Level & $     0.083$ & $     0.207$ & $     0.065$ & $     0.074$ & $     0.068$ & $     0.217$ \\ \cline{1-8}
\multirow{2}{*}{SS} &  $\%$ Flipped    &         N/A  &         N/A  & $   52.73\%$ & $   43.64\%$ & $   52.73\%$ & $   13.64\%$ \\ \cline{2-8}
                    &  Avg. Flip Level &         N/A  &         N/A  & $     0.050$ & $     0.135$ & $     0.050$ & $     0.206$ \\
\bottomrule

\end{tabular}
 \end{table}

\subsubsection{Results}
\label{sec:rq3-results}
Table~\ref{tab:rq3-noise-tolerance} summarizes the results of the noise injection experiments.
The columns correspond to the different types of noise and report the results for both STL* and \lang, while the rows correspond to the different requirement classes.
For each requirement class, we report the percentage of traces whose verdict changes for some noise level (\% Flipped) and the average noise level at which the verdict change occurs (Avg. Flip Level).
In the ``\% Flipped'' row, a lower percentage indicates higher noise tolerance, whereas in the ``Avg. Flip Level'' row a higher average noise level indicates higher noise tolerance.
When no verdict changes are observed, we leave the corresponding cell empty.

We observe that, for the noise types ``Offset'' and ``High-Frequency Noise'', \lang exhibits higher noise tolerance than STL*.
The only exceptions are the SP and OS requirement classes under offset noise; however, the performance difference in these cases is small, and \lang still exhibits a very low percentage of flipped verdicts.
The higher tolerance of \lang to these noise types is related to the ability of the frequency-domain representation to isolate the frequency components that carry information relevant to the system behaviour from the ones that carry the noise.
In particular, a constant offset affects only the low-frequency components, whereas high-frequency noise primarily affects the high-frequency components.
When the specification does not constrain these components, perturbations to them have little or no effect on the monitoring outcome, and hence the frequency-domain specifications are not affected by them.
In contrast, STL* specifications are based on thresholds over signal values at specific points in time and are therefore more sensitive to local changes in the signal induced by noise.

For the white noise experiments, the STL* and \lang specifications compare differently across the different requirement classes.
For SP, ST and SS, \lang consistently outperforms STL*, exhibiting either a lower percentage of flipped verdicts or changing the verdict only for higher levels of noise.
For OS and OV, the performance of the specifications written in the two languages is comparable.
Finally, STL* outperforms \lang for RT, with substantially fewer traces exhibiting a verdict change ($29.23\%$ versus $67.69\%$).
Unlike offset and high-frequency noise, white noise affects the whole frequency spectrum, including the frequency components that encode the relevant system behaviour.
As a result, also \lang specifications cannot effectively filter out its effects, making them more susceptible to this type of noise.

We observe that STL* specifications exhibit nearly identical percentage of flipped verdicts and average flip levels for white and high-frequency noises.
A closer inspection of the data reveals that, in most cases, the same traces experience verdict changes under both types of noise.
This indicates that STL* specifications are equally affected by white and high-frequency noise.
In fact, since STL* specifications rely on signal values at specific points in time, they are influenced by the magnitude of the noise and cannot leverage its frequency characteristics (i.e., correlations between its values over time) to prevent it from affecting the monitoring.
On the contrary, \lang specifications exhibit a substantially higher noise tolerance for traces affected by high-frequency noise, when compared to white noise.
This is because they can exploit the fact that high-frequency noise is confined to a specific region of the spectrum.
Notably, with the high-frequency noise, only RT and SS exhibit more than a few flipped verdicts, with $6.15\%$ and $13.64\%$ of traces affected, respectively.
This showcases the ability of the frequency-domain specifications to isolate the frequency components that encode the relevant system behaviour while ignoring components affected by noise. 
Further, this underlines the ability of the frequency domain representation to analyse the shape of a signal independently of its values at specific points in time.

\begin{summarybox}
    {\bf Summary:} Our experiments show that, when signals are affected by constant offsets or high-frequency noise, the time-frequency-domain specifications are significantly more noise-tolerant than the time-domain ones.
    White noise, however, degrades also the performance of frequency-domain specifications.
    Nevertheless, also with white noise, the frequency domain specifications remain more noise-tolerant than their time-domain counterparts for most requirement classes (SP, ST and SS).
    For OS and OV specifications, the noise tolerance is comparable, whereas, for RT, the time domain performs better.
\end{summarybox}

\subsection{Discussion}
\label{sec:evaluation-discussion}
The motivating hypothesis of this paper is that expressing signal-based CPS requirements using the time-frequency domain representations of signals can improve both the broader applicability of the requirements and the noise tolerance of their monitoring.
The broader applicability is confirmed by the results of RQ1, which
demonstrate the extension of the applicability of requirements expressed in time-frequency domain to input traces using linear or constant-linear alternate interpolation, when compared to the time-domain counterpart.
The noise tolerance of monitoring is instead confirmed by the results of RQ3, which show that the time-frequency specifications are more tolerant to offset and high-frequency noise, while their tolerance to white noise is comparable to that of the time-domain specifications.

These results are complemented by the results of RQ2, which compare the fidelity of time-domain and time-frequency-domain specifications to the intrinsic signal properties.
The RQ2 results show that the relative performance of the two domains varies across different requirement classes.
However, RQ2 considers only traces for which both sets of specifications are applicable.
As discussed in the answer to RQ2 (Section~\ref{sec:rq2-results}), the pure time-domain specifications are more restrictive in their applicability, as shown in RQ1. This allows them to make stronger assumptions about the signal shape, making it easier to achieve higher requirement fidelity (in terms of precision and soundness) in the specification formulation.
In contrast, the time-frequency specifications are applicable also to traces that use linear and constant-linear alternate interpolation, and achieve good requirement fidelity, especially in terms of precision (Table~\ref{tab:rq2-fidelity-extra}).

To conclude, the use of the time-frequency-domain representation of signals enables specifications that are applicable to a broader range of traces and monitoring that is more tolerant to noise with a known spectral profile, as is often the case in practice.
However, pure time-domain specifications remain an effective choice for traces where strong assumptions about the signal shape can be made and the noise level is low.

\subsection{Threats to Validity}
\label{sec:evaluation-threats}
\subsubsection{Internal Validity}
\label{sec:threats-internal}
In terms of internal validity, the main threat is that the STL* specifications were adapted from~\citet{Kapinski:2016b}, and the \lang specifications were written by us using the templates proposed in Section~\ref{sec:templates}.
This poses a validity threat because alternative choices could be made to specify the requirements, potentially leading to different results.
However, both the templates from~\citet{Kapinski:2016b} and our proposed templates are based either directly on the definitions of the requirements themselves or on well-established results from control theory~\cite{Astrom:2008}.
This provides confidence that our formulas are representative among the plausible specification alternatives.
Moreover, the selected parameter values are motivated by engineering considerations, which we documented in the appendices.

Another threat to internal validity arises from the random generation of the noise used in RQ3.
Different sampling of the noise values could have produced different results.
However, our evaluation is based on a relatively large dataset, comprising $1252$ traces in total and at least $65$ traces for each requirement class.
Because each trace was generated using an independently sampled noise sequence, the dataset covers a broad range of possible noise realisations.
Furthermore, when comparing STL* and \lang, we evaluate both specifications on the same noisy trace.
As a result, the comparison is performed under identical conditions, ensuring its fairness.

As noted when reporting for the time-domain formula of the RT requirement (Eq.~\ref{eq:rt-time}), the STL* formula implicitly assumes that the step input starts from the value $0$.
This assumption holds for the step inputs used in RQ2 and RQ3, but does not necessarily hold for the inputs generated using constant interpolation.
When the assumption is violated, the threshold used to evaluate the requirement is slightly lower than intended because the variable $x$ starts from a higher value. 
Consequently, the constraint becomes less restrictive, which may reduce the soundness of the formula, that is, its ability to detect all violations.
Nevertheless, a higher soundness value would not alter our conclusions, as, for the RT, the time-domain representation already outperforms the time-frequency-domain representation.

\subsubsection{External Validity}
\label{sec:threats-external}
The external validity of our experiments depends on the extent to which the selected systems and generated test inputs are representative of real-world CPSs.
For RQ1, we used two real CPSs to determine the parameters for generating the test inputs.
This gives us confidence that the resulting input traces are realistic.
For RQ2 and RQ3, we used instead synthetic systems for the generation of the test outputs.
However, the mathematical model that we used (a second-order linear system) is generally considered a good approximation of a closed-loop control system~\cite{Astrom:2008}.
Therefore, although the systems themselves are not instances of real CPSs, the generated input and output traces can be considered representative of those encountered in practice.

For generating the input traces, we did not employ a specific test generation technique.
Consequently, one could argue that the generated traces may not be representative of actual test cases.
However, we generated the traces by random sampling using the standard parametrisation adopted by many state-of-the-art testing approaches~\cite{Mandrioli:2025, Yadav:2025, Hildebrandt:2020, Matinnejad:2019}.
As such, our traces can be considered representative of the inputs generated by these testing approaches.
Moreover, we evaluated our approach on a large number of traces: $1600$ for RQ1, $1920$ for RQ2, and $1252$ for RQ3.
These sample sizes give us confidence that the observed results are not artefacts of a specific realisation of the random sampling.

Further, for our input generation, we considered only non-negative signal values: step inputs ranged from $0$ to a positive value, and the control points of other input signals were restricted to positive values.
We made this choice because the time-domain specification templates we used assume non-negative signals; otherwise, the inequalities would no longer correctly capture the intended relations.
However, this limitation does not affect our conclusion that the time-frequency-domain representation is a valuable extension of the traditional time-domain approach.
In fact, considering also signals which take negative values would only reduce the applicability, fidelity and noise tolerance of the time-domain formulas.
In contrast, the applicability, fidelity, and noise tolerance of the time-frequency-domain formulas would remain unchanged.
In fact, the sign of a signal affects the phase of its spectrum, but our specification templates predicate only about the amplitude.
Therefore, extending the evaluation to signals with negative values would only strengthen the conclusions drawn from our results.

     \section{Outlook on Software Engineering for CPS with Time-Frequency Specifications} \label{sec:outlook}
In this section, we discuss the implications of the results presented in this paper for both researchers and practitioners.

Our results show that the use of time-frequency signal representations extends the specifications applicability to a broader set of traces and improves the noise tolerance of monitoring.
This has implications for all software engineering (SE) activities
that leverage specification languages and runtime verification~\cite{Bartocci:2018}.
Among the various SE activities, we highlight the potential for new developments in requirements mining, falsification testing, and fault localisation.
For requirements mining, researchers will need to investigate approaches for synthesising the time-frequency transforms used in specifications to capture relevant signal properties~\cite{Nesterini2025}.
This challenge will also be closely intertwined with the learning of specification parameters.
Falsification testing (i.e., the generation of test cases aimed at finding counterexamples to the satisfaction of a given specification) will require new test-generation approaches that explore system behaviours differing in their frequency content rather than merely in their signal values~\cite{Winsten2026}.
Finally, the use of time-frequency representations opens opportunities for developing new fault-localisation approaches based on changes in the spectral properties of signals caused by faults~\cite{Bartocci2022}.

For practitioners, we foresee two main opportunities arising from the use of time-frequency signal representations.
First, traditional software engineering activities can benefit from the results observed in this paper.
Specifically, the increased applicability and noise tolerance of monitoring can make specification development and maintenance~\cite{Fennell2020,Bowen1993} more cost-effective, as specifications can be used as oracles on a larger number of traces.
Second, the use of a specification language based on this signal representation can simplify interactions among software engineers, systems engineers, and control engineers~\cite{mandrioli:2023b}.
Indeed, the latter often work with frequency-domain representations of systems and signals.
A formalism capable of expressing both time-domain and frequency-domain properties can provide common ground for software engineers to collaborate more effectively with experts from other disciplines.
In conclusion, these opportunities make the industrial adoption~\cite{Dawes2024} of formal specifications and runtime verification techniques more likely.

     \section{Related Work}
\label{sec:related}

\subsection{Time-Based Specification Languages}
\label{sec:related-time-domain}
Researchers have proposed a variety of languages for specifying signal-based CPS requirements.
Arguably, most state-of-the-art languages originate from Linear Temporal Logic (LTL)~\cite{Pnueli:1977}, which enables the expression of constraints over the temporal ordering of propositional variables, and its extension, Metric Temporal Logic (MTL)~\cite{Koymans:1990}, which augments LTL with quantitative timing constraints. 
Motivated by applications to control systems and analog and mixed-signals circuits, MTL was further extended to Signal Temporal Logic (STL)~\cite{Maler:2004}, which enables expressing constraints over dense-timed, real-valued signals.\footnote{
    More precisely, STL extends Metric Interval Temporal Logic (MITL), which is a decidable fragment of MTL.
}
At its core, STL introduces atomic propositions that express inequality constraints over signal values.
Notably, STL is the specification language adopted by~\citet{Kapinski:2016b} to define templates for the specification classes in their taxonomy (discussed in Section~\ref{sec:setup-langs}).
Despite its widespread adoption, researchers have highlighted several limitations of STL and have proposed numerous extensions.
Based on two surveys~\cite{Bartocci:2018,Chaima:2021}, we summarize the main extensions to STL and refer the reader to the original surveys for a more comprehensive overview and discussion.

\partparagraph{STL*}
STL can only compare signals with constants or with other signals at the same time.
As a result, it cannot express comparisons between a signal's values at different points in time.
For example, specifying a local maximum requires comparing the signal's value at a given time instant with its values over adjacent time intervals.
To address this limitation, \citet{Brim:2014} introduced STL*, which enables such comparisons through a ``signal-value-freeze'' operator.
The authors highlight its usefulness for specifying local minima and maxima, which in turn can be used to express properties such as increasing or decreasing oscillations.

\partparagraph{HLS and SCSL}
Hybrid Logic of Signals (HLS) is another language that extends STL~\cite{Menghi:2021}.
Its objective is to facilitate the specification of constraints over both software and physical behaviours.
To this end, HLS introduces quantification over both signal time stamps and time indexes, enabling more flexible reasoning about the temporal dimension.
Later, HLS was integrated with Inter-procedural Control-Flow Temporal Logic (iCFTL)~\cite{dawes2021} into the new language Source Code and Signal Logic (SCSL)~\cite{Dawes:2022}.
This integration enables the combination of source-code level specifications from iCFTL with signal-based specifications from HLS.

Besides the ones discussed above, we also mention three additional specification languages suitable for expressing signal-based CPS requirements that do not extend STL.

\partparagraph{SCL}
\citet{Silvetti:2018} introduced Signal Convolution Logic (SCL) to specify properties that require a predicate to hold for only a given percentage of a time interval, while also enabling the filtering of noise.
This is achieved through a convolution operator that convolves the signal with a kernel (allowing noise filtering) and evaluates whether the predicate holds for the required fraction of the interval.

\partparagraph{SFO}
\citet{Bakhirkin:2018} introduced Signal First-Order Logic (SFO), which allows the use of quantifiers (existential and universal) over both value and time variables.
This overcomes the limitation of STL that it requires comparing signals to known values (i.e., constants or other signals' values).
The authors highlight that the quantification over time variables eliminates the need for temporal operators like ``finally'', ``globally'', and ``until''.

\partparagraph{Shape Expressions}
To enable the description of signal shapes and their detection in signals, \citet{nickovic:2019} proposed Shape Expressions.
In shape expressions, signal shapes are described as sequences of basic patterns, such as linear increase, exponential growth, or sinusoidal oscillations.
These sequences can be written using a grammar similar to that of regular expressions.

In terms of tools for runtime verification, to the best of our knowledge, among the works reported here, RTAMT~\cite{Nickovic:2020,yamaguchi2024} (the one that we used in Section~\ref{sec:evaluation} to build an \lang monitor) supports STL and its extension STL*.
The authors of HLS and SCSL provide replication packages containing monitors for their proposed specification languages.
All these tools are available under permissive open-source licenses.

All the languages discussed above essentially address the common challenge of comparing the values of the same signal at different points in time.
This capability is fundamental for describing signal shapes, which in turn are essential for expressing many CPS requirements.
In this work, we address the same challenge, but take a fundamentally different approach by leveraging the time-frequency representation of signals rather than relying solely on their time-domain representation.

\subsection{Use of the Frequency Domain for Signal-Based CPS Specifications}
\label{sec:related-frequency-domain}
\citet{Chakarov:2012} were the first to propose the use of the frequency domain for specifying signal-based CPS requirements.
In this preliminary work, the authors proposed to define constraints on the amplitude components of a signal's frequency spectrum.
They also addressed the signal generation problem by proposing an approach to generate signals that satisfy a given specification.
The main limitation of this work is that it considers only periodic signals and relies exclusively on the Fourier transform, thereby completely discarding the temporal dimension of the signal.

\partparagraph{TFL}
\citet{donze:2012} extended STL with the Short-Time Fourier Transform to introduce Time-Frequency Logic (TFL).
Their objective was to handle non-periodic signals by enabling predicates over how a signal's frequency spectrum evolves over time.
Specifically, they extended STL with inequality constraints on the amplitude of individual frequency components.
As a proof of concept, they applied TFL to automatically recognise a blues melody in a song.
Subsequently, \citet{Nguyen:2017} enhanced TFL by replacing STFT with
the Wavelet transform and defining a parametric version of the
language (called Parametric Time Frequency Logic - PTFL), which allows for unknown parameters in the formulas that can be learned from traces rather than being set by the user.
They used PTFL for anomaly detection in a hydrogen fuel cell and an electrocardiogram.
Later, also \citet{Beg:2021} used PTFL for anomaly detection in electric microgrids.
To the best of our knowledge, neither an implementation of TFL nor PTFL are available.

Our work is closely related to TFL, as \lang also leverages the time-frequency representation of signals.
However, \lang places a stronger emphasis on the frequency domain by supporting more expressive predicates over the time-frequency spectrum of one or more signals (Eq.~\ref{eq:lang-fd-predicates}), as well as predicates over frequency intervals rather than individual frequency components.
As discussed in Section~\ref{sec:desiderata}, these features are essential for specifying CPS requirements.

Finally, we mention a few additional works that leverage the frequency-domain representation of signals in the context of CPS V\&V.
\citet{rodinova:2016} proposed continuous-time LTI filters that implement the semantics of MTL.
\citet{Basnet:2020} performed a Fourier analysis of STL monitors.
\citet{mandrioli:2023b} and \citet{Yadav:2025} exploited the frequency domain for test case generation.
While these works do not directly address the specification of CPS requirements, they underline the relevance and potential of the frequency domain for the V\&V of CPSs.

     \section{Conclusions}
\label{sec:conclusions}
In this paper, we have investigated the use of time-frequency representation of signals for specifying signal-based CPS requirements.
We have analysed existing taxonomies of CPS requirements and identified the requirement classes that can benefit from time-frequency representations.
Based on this analysis, we have derived the desiderata for a specification language operating on time-frequency representations and proposed \langFull, a logic that enables predicates over frequency intervals as well as logical relations between frequency components.
We have further proposed specification templates for the identified requirement classes and used them to empirically compare the formulation of requirements in the time and time-frequency domains.
Our empirical results show that the time-frequency representation substantially broaden the applicability of step-response requirements by eliminating their dependence on step-like inputs and enabling their evaluation on traces generated through linear interpolation.
Moreover, requirements specified in the time-frequency domain exhibit increased tolerance to signal offsets and high-frequency noise while maintaining comparable fidelity in capturing the intended signal properties.
Overall, our findings demonstrate that the time-frequency representations provide a valuable extension to the traditional time-domain representations for specifying CPS requirements, particularly by broadening the applicability and improving the noise tolerance of monitoring.

In the future, we plan to investigate the practical aspects of using \lang for specifying CPS requirements.
Specifically, we will conduct a user study to understand its potential for adoption by software engineering practitioners in the field of CPS, investigating the understandability of \lang specifications~\cite{Czepa2019,Czepa2020}.
At the same time, we plan to develop GUI-based tools to facilitate the use of \lang, for example with graphical representations of the formulas in relation to the monitored signals.
On the more theoretical side, we will investigate the possibility of implementing online monitors, in light of the discussion in Section~\ref{sec:lang-implementation}.

    \section*{Acknowledgments}
        This work has received funding from the European Union’s Marie Skłodowska-Curie Actions – Postdoctoral Fellowships (PF) program under Grant Agreements No.\ 101148870 (ContTestCPS) and No.\ 101273739 (REACT-CPS).
        For the purpose of open access, and in fulfilment of the obligations arising from the grant agreements, the authors have applied a Creative Commons Attribution 4.0 International (CC BY 4.0) license to any Author Accepted Manuscript version arising from this submission.

    \bibliographystyle{plainnat}

    \clearpage
    \appendix
    \section{Setting of Specifications Parameters}
\label{sec:setting-specs-params}
\subsection{RQ1}
\label{sec:app-rq1-params}
Based on the observed step responses, the bandwidth $f_b$ is approximately \qty{1}{\hertz} for the drone and \qty{0.2}{\hertz} for the aircraft.
Intuitively, these values correspond to reaction times on the order of seconds for the drone and tens of seconds for the aircraft.
Table~\ref{tab:rq1-parameters} reports the parameter values used in the precondition formulas for RQ1 (Section~\ref{sec:rq1}), together with the rationale for their selection.
\begin{table}[h]
    \centering
    \caption{Parameter values used in the precondition specifications. The ``f.int.'' rows indicate the frequency intervals selected for the frequency predicates.}
    \label{tab:rq1-parameters}
    \small
    \begin{tabular}{lllp{0.5\textwidth}}
\toprule

\textbf{Parameter}             & \textbf{Drone}                & \textbf{Aircraft}          & \textbf{Explanation} \\ \midrule

$t_{\max}$                     & \qty{15}{\second}             & \qty{120}{\second}         & The drone completes manoeuvres in a few seconds; the aircraft in tens of seconds. \\
$\sigma$                       & \qty{0.1}{\metre}             & \qty{50}{\metre}           & A relevant input change for the drone is \qty{10}{\centi\metre} and for the aircraft is \qty{50}{\metre}. \\
$\tau$                         & \qty{3}{\second}              & \qty{9}{\second}           & Set equal to the settling time. \\
$\varepsilon$                  & \qty{5}{\milli\metre/\second} & \qty{0.25}{\metre/\second} & A rate of change below a few \unit{\milli\metre} for the drone and a few \unit{\deci\metre} for the aircraft practically corresponds to a constant signal. \\
$\mathit{ss}_{\mathit{th}}$    & \qty{5}{\milli\metre}         & \qty{0.05}{\metre}         & Similar to $\varepsilon$. For the aircraft, a smaller threshold can be used because the signals are more stable. \\
SS f.int.                     & $[0.5,2]\unit{\hertz}$        & $[0.1,1]\unit{\hertz}$     & Frequency range around and above $f_b$. \\
$\mathit{ov}_{\mathit{th}}$    & \qty{5}{\milli\metre}         & \qty{6}{\metre}            & Chosen to capture relevant input changes, but smaller than $\sigma$ because these correspond to high-frequency components. \\
OV f.int.                      & $[0.25,2]\unit{\hertz}$       & $[0.05,1]\unit{\hertz}$    & Frequency range around $f_b$. Wider than SS on the lower bound to better capture possible amplifications. \\
$\mathit{rt}_{\mathit{th}}$    & \qty{5}{\milli\metre}         & \qty{0.05}{\metre}         & Same as $\mathit{ss}_{\mathit{th}}$. \\
RT f.int.                      & $[0.8,3]\unit{\hertz}$        & $[1.15,4]\unit{\hertz}$    & Frequency range above $f_b$. \\
$\mathit{ST}$                  & \qty{3}{\second}              & \qty{9}{\second}           & Chosen based on the observed convergence time of the step responses. \\
$\mathit{ss}_{\mathit{th}\_r}$ & \qty{5}{\milli\metre}         & \qty{0.05}{\metre}         & Same as $\mathit{ss}_{\mathit{th}}$. \\
ST f.int.                      & $[0.5,2]\unit{\hertz}$        & $[0.1,1]\unit{\hertz}$     & Same frequency interval as for SS. \\

\bottomrule

\end{tabular}
 \end{table}

\subsection{RQ2}
\label{sec:app-rq2-params}
All formulas begin with either a $\square$ or $\Diamond$ operator over a time interval covering the entire trace: \qty{20}{\second} for SP and OS, and \qty{100}{\second} for RT, ST, OV and SS.
In the templates presented in Section~\ref{sec:templates}, we used $f_{\max}$ to denote the maximum frequency of the time-frequency representation.
However, in practice, most high-frequency components have very small amplitude and carry little or no useful information.
In our experiments, $f_b$ is typically around \qty{1}{\hertz}, with the relevant frequency components lying below \qty{3}{\hertz}.
Therefore, in all formulas we set $f_{\max}$ to \qty{4}{\hertz}, avoiding unnecessary evaluation of very high-frequency components that do not carry useful information.

\subsubsection{SP}
\begin{itemize}
    \item [{\bf STL}:] The value $0.01$ used to compute the discrete derivative corresponds to the sampling time.
    The \qty{2}{\second} time bound of the $\Diamond$ operator represents an upper bound on the spike width.
    The amplitude threshold of $5$ is adopted from the original work~\cite{Kapinski:2016b}, which assumes an expected spike amplitude of $10$.
    Specifically, the threshold is chosen such that the product of the spike amplitude and the derivative threshold remains proportional to the expected amplitude.
    \item [{\bf \lang}:] The \qty{1}{\second} time bound of the $\square$ operator represents a lower bound on the spike width.
    The threshold value of $1$ for the time-frequency transform is derived from the actual expected spike amplitude of $10$, but is reduced by one order of magnitude as it is a higher-frequency component.
    The frequency interval $[0.5,4]$ is chosen to include components above \qty{0.5}{\hertz}, as those include the expected longest spike duration of \qty{2}{\second}\textemdash in fact, the duration is inverse to the frequency.
\end{itemize}

\subsubsection{OS}
\begin{itemize}
    \item [{\bf STL}:] The amplitude thresholds of $20$ and $40$ are based on the expected peak-to-peak amplitude (i.e., twice the expected amplitude range of $10$ to $20$).
    The time interval of the $\Diamond$ operator is set to $[0.48,0.52]\unit{\second}$, providing a small margin around half of the expected oscillation period.
    The time shift of the $\ast$ operator is set to \qty{0.25}{\second}, corresponding to half of the time interval of the $\Diamond$ operator, as prescribed in the original work~\cite{Kapinski:2016b}.
    \item [{\bf \lang}:] The amplitude thresholds of $4.5$ and $10$ are set to approximately half of the expected amplitude of the oscillations to account for the scaling of the time-frequency representation.
    The frequency interval $[0.85,1.15]\unit{\hertz}$ is chosen to capture the expected oscillation period of \qty{1}{\second}, corresponding to a frequency of \qty{1}{\hertz}.
    The frequency intervals $[0.4,0.85]\unit{\hertz}$ and $[1.15,1.6]\unit{\hertz}$ are used to characterize the low-amplitude frequency intervals around the frequency range where we expect to see the oscillation.
\end{itemize}

\subsubsection{RT}
\begin{itemize}
    \item [{\bf STL}:] The value of $1$ specifies the minimum size of the considered step changes in the input.
    The time interval of the $\Diamond$ operator is set to $[0,0.75]\unit{\second}$, corresponding to the prescribed $0\%$ to $75\%$ rise time reported in Table~\ref{tab:rq2-sys-values}.
    \item [{\bf \lang}:] The prescribed $0\%$ to $75\%$ rise time of \qty{0.75}{\second} corresponds to a bandwidth of approximately \qty{0.3}{\hertz}.
    Accordingly, for the low-frequency part of the specification we use a predicate over the interval $[0,0.3]\unit{\hertz}$, while the high-frequency part considers frequencies above $[0.6]\unit{\hertz}$.
    The value of $0.25$ for the low-frequency error is derived from the value for the SS requirement, but is increased because the RT predicates about the transients of the output tracking the input, where a larger tracking error is expected.
    Since the input amplitude can reach $10$, a threshold of $0.25$ corresponds to a maximum error of $2.5\%$.
    For the high-frequency part of the specification, which checks for filtering behaviour, we use a threshold of $0.1$. Because the input amplitude can be as high as $10$, this threshold excludes time-frequency components whose amplitudes are too small to support a reliable assessment.
\end{itemize}

\subsubsection{ST}
\begin{itemize}
    \item [{\bf STL}:] The value of $1$ specifies the minimum amplitude of the considered step changes in the input.
    The time interval $[4.5,6]\unit{\second}$ in the $\square$ operator is based on the prescribed settling time of \qty{4.5}{\second}, allowing for an additional \qty{1.5}{\second} window during which the input-output difference is evaluated.
    The coefficient $0.02$ is based on the $2\%$-range definition of the settling time.
    \item [{\bf \lang}:] The upper bound of the time interval $[0,2.25]$ in the until operator is chosen based on the prescribed settling time of \qty{4.5}{\second}.
    We use half of the actual settling time requirement because the time-frequency transform reduces the resolution along the time axis, and thus requires additional slack.
    Regarding the frequency intervals, the settling time of \qty{4.5}{\second} corresponds to an approximate bandwidth of \qty{0.15}{\hertz}.
    Thus, to check that the input and output are at steady state, we predicate over the frequency range $[0.15,4]$\unit{\hertz}.
    The threshold of $0.1$ is based on the minimum relevant unit changes, but reduced by one order of magnitude because the predicate operates on higher-frequency components.
\end{itemize}

\subsubsection{OV}
\begin{itemize}
    \item [{\bf STL}:] The value of $1$ specifies the minimum magnitude of the considered step changes in the input.
    The time interval of the $\square$ operator is set to $[0,2]\unit{\second}$ to fully capture the transient during which the output tracks an input change.
    The coefficient $0.15$ is based on the maximum allowable overshoot (Table~\ref{tab:rq2-sys-values}).
    \item [{\bf \lang}:] The predicate is evaluated over the frequency range $[1,3]\unit{\hertz}$, corresponding to the system bandwidth of \qty{2}{\hertz}.
    Because the input and output changes are in the range of units while the predicate targets high-frequency components, we use a threshold of $0.05$ to disregard insignificant output variations.
    The coefficient of $1.15$ implements the allowable $15\%$ overshoot by permitting an output amplification of up to (15\%) relative to the input.
\end{itemize}

\subsubsection{SS}
\begin{itemize}
    \item [{\bf STL}:] To detect steady state, we select a threshold of $0.01$ to the discrete derivative, which corresponds to a rate of change of $1$ unit per second (analogous to the minimum input step of size $1$ considered in the other specifications).
    Following the original formulation in~\cite{Kapinski:2016b}, for the until operator, we consider the time interval $[2,2.02]\unit{\second}$ to evaluate the tracking error after the transient, as the generated system has a settling time of \qty{2}{\second}.
    The value of $0.1$ corresponds to the maximum allowable static error specified in Table~\ref{tab:rq2-sys-values}.
    \item [{\bf \lang}:] Since the settling time of \qty{0.75}{\second} corresponds to a frequency bandwidth of \qty{1}{\hertz}, to detect the steady-state behaviour of the input, we consider the frequency range $[0.5,4]\unit{\hertz}$.
    The upper threshold on the high-frequency components of the input is set to $0.01$, based on the allowable static error, but reduced because the predicate operates on high-frequency components.
    The time-frequency spectrum of the tracking error is bounded by $0.1$, in accordance with the specification (Table~\ref{tab:rq2-sys-values}), and is evaluated over the low-frequency range of $[0,0.2]\unit{\hertz}$.
\end{itemize}
 
\end{document}